\documentclass[journal]{IEEEtran}

\IEEEoverridecommandlockouts
\usepackage{cite}
\usepackage{amsmath,amssymb,amsfonts}
\usepackage{algorithmic} 
\usepackage{algorithmic,algorithm}
\usepackage{graphicx}
\usepackage{textcomp}
\usepackage{xcolor}
\usepackage{algorithm}
\usepackage{amsthm}
\usepackage{enumerate} 
\theoremstyle{plain}
\theoremstyle{definition}
\theoremstyle{remark}

\usepackage{multirow}  
\def\BibTeX{{\rm B\kern-.05em{\sc i\kern-.025em b}\kern-.08em
		T\kern-.1667em\lower.7ex\hbox{E}\kern-.125emX}}

\newtheorem{theorem}{\textbf{Theorem}}

\newtheorem{lemma}{\textbf{Lemma}}

\newtheorem{corollary}{Corollary}

\newtheorem{definition}{\textbf{Definition}}
\newcommand{\tabincell}[2]{\begin{tabular}{@{}#1@{}}#2\end{tabular}}

\usepackage{graphicx}
\usepackage{subfigure} 
\ifCLASSOPTIONcompsoc
\usepackage[caption=false,font=normalsize,labelfont=sf,textfont=sf]{subfig}
\else
\usepackage[caption=false,font=footnotesize]{subfig}
\fi

\makeatletter

\newcommand{\Rmnum}[1]{\expandafter\@slowromancap\romannumeral #1@}
\makeatother

\usepackage{flushend}

\begin{document}
	
\bstctlcite{ref:BSTcontrol}

\title{Online Offloading Scheduling for NOMA-Aided MEC Under Partial Device Knowledge}

\author{Meihui~Hua,
	Hui~Tian,~\IEEEmembership{Senior Member,~IEEE,}
	Xinchen~Lyu,~\IEEEmembership{Member,~IEEE,} \\
	Wanli~Ni,~\IEEEmembership{Student~Member,~IEEE,}
	and Gaofeng~Nie~\IEEEmembership{Member,~IEEE}
	\thanks{This work was supported by the National Nature Science Foundation of China under Grant No. 61801046. A short version of this paper was presented at IEEE ICC Workshops, Virtual, Montreal, Canada, Jun. 2021 \cite{Hua2021Online}. \textit{(Corresponding author: Hui Tian.)}}
	\thanks{M. Hua, H. Tian, W. Ni, and G. Nie are with the State Key Laboratory of Networking and Switching Technology, Beijing University of Posts and Telecommunications, Beijing, China (e-mail: huameihui@bupt.edu.cn; tianhui@bupt.edu.cn; charleswall@bupt.edu.cn; niegaofeng@bupt.edu.cn).}
	\thanks{X. Lyu is with the National Engineering Laboratory for Mobile Network Technologies, Beijing University of Posts and Telecommunications, Beijing, China (e-mail: lvxinchen@bupt.edu.cn).}
	\thanks{Copyright (c) 2021 IEEE. Personal use of this material is permitted. However, permission to use this material for any other purposes must be obtained from the IEEE by sending a request to pubs-permissions@ieee.org.}
	}

\maketitle

\begin{abstract}
	By exploiting the superiority of non-orthogonal	multiple access (NOMA), NOMA-aided mobile edge computing (MEC) can provide scalable and low-latency computing services for the Internet of Things. However, given the prevalent stochasticity of wireless networks and sophisticated signal processing of NOMA, it is critical but challenging to design an efficient task offloading algorithm for NOMA-aided MEC, especially under a large number of devices. This paper presents an online algorithm that jointly optimizes offloading decisions and resource allocation to maximize the long-term system utility (i.e., a measure of throughput and fairness). Since the optimization variables are temporary coupled, we first apply Lyapunov technique to decouple the long-term stochastic optimization into a series of per-slot deterministic subproblems, which does not require any prior knowledge of network dynamics. Second, we propose to transform the non-convex per-slot subproblem of optimizing NOMA power allocation equivalently to a convex form by introducing a set of auxiliary variables, whereby the time-complexity is reduced from the exponential complexity to $\mathcal{O} (M^{3/2})$. The proposed algorithm is proved to be asymptotically optimal, even under partial knowledge of the device states at the base station. Simulation results validate the superiority of the proposed algorithm in terms of system utility, stability improvement, and the overhead reduction.
\end{abstract}

\begin{IEEEkeywords}
Mobile edge computing, offloading schedule, stochastic optimization, non-orthogonal multiple access
\end{IEEEkeywords}

\section{Introduction}
		The flourish of innovative applications in the Internet of Things (IoT), such as intelligent transportation, navigation, environmental monitoring, brings the explosive expansion of mobile data traffic in wireless networks \cite{Aazam2018Deploying,Wang2019MOERA}. 
		The tremendous growth of data can overburden resource-constrained IoT devices and lead to the increase of network overhead and latency in conventional cellular networks \cite{Liang2020Toward}. To cope with the unprecedented scale of communication and the emergence of resource-demanding computational requests, mobile edge computing (MEC) is concerned as a promising and realistic approach to upgrade the IoT systems  \cite{Lyu2017Multiuser}. The fundamental insight behind MEC is to deploy communication, computation, control, and caching capacities in close proximity to mobile terminals. In other words, edge facilities, such as base stations (BS), can dominate partial-sized cloud-like infrastructure distributed across the network and provide computation offloading services to nearby devices. By deploying a large amount of sensors to monitor the time-varying status of the physical world, the ubiquitous sensors in IoT networks will generate massive real-time data. These data need to be processed in time with the aim of making judicious decisions to control the reaction of IoT devices to the physical world \cite{Lei2020Deep}. However, since IoT devices always have limited computational and memory capacity, it is impractical to process and storage all these data locally. By virtue of the MEC, millions of IoT devices are allowed to offload computation tasks to the capable edge servers. The performance of IoT devices can be greatly improved by offloading complex tasks to capable edge servers. The trend towards MEC is expected to accelerate as more devices perform a majority of computations and as more intensive applications get adopted. However, offloading from massive IoT devices intensifies the network link burden and it presents requirement on the high throughput of the MEC network \cite{Xu2020Fairness,Li2020Throughput,Yang2020On}. The extremely limited and costly spectrum resources become the bottleneck of guarantee the quality of network service \cite{Premsankar2018Edge,Fan2018Application}.
		
		In the traditional orthogonal multiple access (OMA) network, each user is allocated a mutually orthogonal resource blocks, such as unique time slot and orthogonal subcarrier. As the number of IoT devices increases, OMA technology is approaching the fundamental limits in throughput and access capacity due to the inherent scarcity of orthogonal resources. Non-orthogonal multiple access (NOMA) is a potential candidate for the air interface techniques of IoT system. Power domain NOMA serves multiple users simultaneously over the same physical resource block (PRB) via the power-domain division, and applies multiuser detection algorithms such as successive interference cancellation (SIC) for inter-user interference cancellation at the receiver side. NOMA has the potential to accommodate more users over the limited number of available orthogonal communication resources \cite{Pham2020Coalitional}. It has been verified in both theory \cite{Dai2015Non} and system-level simulations \cite{Saito2013Non} that NOMA outperforms orthogonal frequency-division multiple access (OFDMA) in terms of spectrum efficiency and device connections. However, power allocation is an essential issue in NOMA, since the co-channel interference is artificially induced among users. Besides, considering the disparities between power limitation of IoT devices and the stringent computation requirements, offloading decisions and computational resource of devices are also required to be optimized properly. Therefore, from the perspective of IoT users, offloading scheduling including task offloading decisions, local computational resource allocation, and offloading power control are critical issues to achieve the full benefit of MEC and NOMA \cite{Zhu2017On}.

\subsection{Challenges}
	Inspired by the aforementioned backgrounds, we are interested in investigating the throughput improvement by integrating NOMA and MEC. To this end, offloading scheduling should be designed properly to jointly determine the offloading decisions, computational resource allocation and power control along the time. However, offloading scheduling in the considered NOMA-aided MEC system confronts with the following technical challenges.
\begin{itemize}
	\item  \textbf{Complexity of NOMA}. 
	In NOMA, the receiver is required to successively decode the signals of the devices (using the PRB) in turn (i.e., the process of SIC) to obtain their payloads. The complexity of SIC would increase significantly with the number of devices. As a result, the user scheduling and power control algorithm need to be meticulously designed for the effectiveness of NOMA-aided MEC.
	
	\item  \textbf{Network Stochasticity}.  
	In the presence of the randomness of data arrival, stochastic wireless channels, and sophisticated availability of resources in realistic wireless networks, it is challenging to make efficient and online scheduling decisions without a priori knowledge of network dynamics. Moreover, the decisions of task offloading and resource scheduling are coupled in the time domain, and myopic optimization would result in significant performance degradation. Making optimal offloading decisions requires statistical information of the involved stochastic processes, and should strike a good balance between the system performance of the current and future features. 
	
	\item  {\textbf{Large-Scale Networking}}. 
	First, the offloading transmission in large-scale networks becomes a multi-objective optimization problem, which gets complicated as the network size scales up. Second, transmitting tremendous data simultaneously is more likely to cause severe access collision, and massive communication access at the BS incur a substantial signaling overhead, which may suffer from the curse of dimensionality. As a consequence, offloading strategies to attain collaborative behavior may reasonably be developed by having the BS relying only on a partial knowledge of the overall state of the network. 
	
\end{itemize}

\subsection{Related Work}

\begin{table*}[t] 
	\centering
	\caption{Summary of related work on NOMA-aided MEC networks}
	\label{Related_Work}
	\begin{tabular}{|c|c|c|c|c|c|}
		\hline
		\textbf{References} & \textbf{Optimization objective}  & \textbf{Stochasticity} & \tabincell{cc}{{\textbf{Partial device}}\\{\textbf{knowledge}}}  & \textbf{Methodology} & \textbf{Optimality} \\
		\hline
		\cite{Kiani2018Edge} & \multirow{3}*{Energy consumption minimization }& $\times$ & $\times$ & Heuristic algorithm & Suboptimal \\
		\cline{3-6}
		\cite{Ding2019Joint} & & $\times$ & $\times$ & Convex optimization & Optimal \\
		\cline{3-6}
		\cite{Fang2019Optimal} & &$\times$ & $\times$ &  Bilevel programming & Optimal \\
		\hline
		\cite{Huang2020NOMA} & Transmission rate maximization & $\times$ & $\times$ & Convex optimization & Optimal \\
		\hline
		\cite{Wu2018NOMA} & \multirow{2}*{Delay minimization} & $\times$ & $\times$ & Distributed algorithm & Optimal \\
		\cline{3-6}
		\cite{Qian2020Latency} &  & $\times$ & $\times$ & Layered algorithm & Optimal \\
		\hline
		\cite{Liu2020SWIPT} & Transmit power minimization & $\times$ & $\surd$ & Convex optimization & Suboptimal\\
		\hline
		\cite{Gong2018Antenna} & \multirow{2}*{Ergodic rate and outage probability optimization} & $\times$ & $\surd$ & Algebraic method & N.A.\\
		\cline{3-6}
		\cite{Wan2018Cooperative} & & $\times$ & $\surd$ & Algebraic method & N.A.\\
		\hline
		\cite{Cao2019A} & Energy efficiency maximization & $\surd$ & $\times$ & \tabincell{cc}{Game-theoretic\\algorithm} & Suboptimal \\
		\hline
		\cite{Yang2018Deep} & Average end-to-end reliability maximization & $\surd$ & $\times$ & {Deep Q-learning} & Optimal \\
		\hline
		\cite{Wang2019Task} & Delay and energy consumption minimization & $\surd$ & $\times$ & Deep Q-learning & Suboptimal \\
		\hline
		\cite{Qian2020NOMATII} & Energy consumption minimization & $\surd$ & $\times$ & \tabincell{cc}{Deep reinforcement\\learning} & Suboptimal \\
		\hline
		\cite{Tuong2020Delay} & Delay minimization & $\surd$ & $\times$ & Deep Q-learning & Optimal \\
		\hline
		\cite{Wang2019Cooperative} & Energy consumption minimization & $\surd$ & $\times$ & Lyapunov optimization & \tabincell{cc}{Asymptotically\\optimal} \\
		\hline
		\cite{Zhang2020Dynamic} & Enery efficiency maximization & $\surd$ & $\times$ & Lyapunov optimization & \tabincell{cc}{Asymptotically\\optimal} \\
		\hline
		\cite{Wang2019User} & \multirow{2}*{{Sum rate maximization}}& $\surd$ & $\times$ & Coalitional game & Locally optimal \\
		\cline{3-6}
		\cite{Zhu2019Optimal} & & $\surd$ & $\times$ & Convex optimization & Optimal \\
		\hline
		\cite{Cao2019Online} & Throughput and fairness & $\surd$ & $\times$ & Lyapunov optimization & \tabincell{cc}{Asymptotically\\optimal} \\
		\hline
		\cite{Nouri2020Dynamic} & Power consumption minimization & $\surd$ & $\times$ & Lyapunov optimization & \tabincell{cc}{Asymptotically\\optimal} \\
		\hline
		This paper & Long-term throughput and fairness maximization & $\surd$ & $\surd$ & Lyapunov optimization & \tabincell{cc}{Asymptotically\\optimal} \\
		\hline
	\end{tabular}
\end{table*}

	Various NOMA-aided MEC schemes have been proposed by academia and industry in recent years, and are briefly summarized in Table \ref{Related_Work}. One-slot optimization of task offloading for NOMA-aided MEC has been studied in \cite{Kiani2018Edge,Ding2019Joint,Fang2019Optimal,Huang2020NOMA,Qian2020Latency,Liu2020SWIPT,Wu2018NOMA,Gong2018Antenna,Wan2018Cooperative} to enhance system	performance in different aspects. To minimize the energy consumption, a heuristic algorithm was designed in \cite{Kiani2018Edge} by jointly optimizing user clustering, computing and communication resource allocation as well as power control. Offloading mode selection was further taken into consideration in \cite{Ding2019Joint} and \cite{Fang2019Optimal}. The transmission rate was maximized in \cite{Huang2020NOMA} by cooperatively offloading data to a helper and an AP using NOMA. Considering the delay minimization problem, optimal offloading solutions were proposed in \cite{Wu2018NOMA} in both single-user and multiple-user cases. User pairing, offloading duration and computation resource allocation were jointly optimized in \cite{Qian2020Latency}. The ergodic rate and outage probability of NOMA were proved to be superior to OMA in both the downlink multiple-input multiple-output NOMA systems with partial channel state information \cite{Gong2018Antenna} and the NOMA-based cooperative relay system with statistical channel state information \cite{Wan2018Cooperative}. However, the aforementioned offline algorithm cannot accommodate the realistic networks primely over a longer time horizon due to the time-variant and stochastic wireless mobile environment. To this end, some recent works have investigated online optimization incorporating the characteristics of the unpredictability of networks. 	
	
	Game-theoretic and deep learning (DL) approaches are popular data-driven dynamic modeling techniques, which are widely applied in stochastic optimization problems. In \cite{Cao2019A}, the energy efficiency maximization problem is modeled as a Stackelberg game, where the subchannel reuse assignment is solved by employing one-to-many matching and the power allocation is tackled by noncooperative game algorithm. 	In \cite{Yang2018Deep}, a real-time adaptive policy based on DL for computational resource allocation is conducted to improve the average end-to-end reliability. A deep Q-learning based algorithm was designed in \cite{Wang2019Task} to achieve delay and energy consumption minimization. Based on deep reinforcement learning, the dalay and total energy consumption were minimized in \cite{Qian2020NOMATII} and \cite{Tuong2020Delay}, respectively, to efficiently learn the near-optimal offloading solutions for the time-varying channel realizations. Nonetheless, game-theoretic	approaches can result in optimality loss due to myopic schedule. The DL models sustain high complexity, since obtaining an accurate model requires sufficient training data set, which raises new challenges to capacity-limited edge facilities. Besides, DL models are qualified as a black box with poor interpretability, whose performance is difficult to quantize mathematically.
	
	Apart from the dynamic algorithms mentioned above, Lyapunov optimization is a powerful technique to analyze stochastic networks with time variations and ensure different forms of system stability. A long-term energy consumption minimization problem was considered in a cooperative MEC system in \cite{Wang2019Cooperative}, where offloading can occur not only between users and MEC servers, but also among MEC servers. In order to address the energy efficiency problem with stochastic task arrivals and time-varying channel states, a dynamic offloading and resource management strategy was designed in \cite{Zhang2020Dynamic}. However, OFDMA was assumed to be used in uplink transmission, which circumvented the inter-user interference but degraded the spectrum efficiency. By contrast, a cooperative NOMA with MEC was investigated in \cite{Wang2019User,Zhu2019Optimal} to show that, with proper power control and offloading strategy, NOMA-MEC system can outperform OMA counterpart in terms of achievable sum rate. The long-term average network utility maximization problem was investigated for a buffer-aided cooperative NOMA system in \cite{Cao2019Online}, where a source was connected to multiple destinations via a buffer-aided NOMA relay. An average weighted-sum power consumption minimization problem was investigated for a stochastic NOMA-based MEC network in \cite{Nouri2020Dynamic}. However, the offloading scheduling in the above works were optimized with the assumption of complete knowledge of the overall users, such as channel state information, computational task arrivals and memory management status, which may not be practical in realistic situations. Since the Lyapunov optimization constructs an approximately optimal solution, studying the performance loss caused by model approximation and incomplete knowledge is of theoretical and practical significance.

\subsection{Contributions}
	Different from the state of the art, this paper proposes a novel asymptotically optimal online offloading scheduling scheme in the NOMA-aided MEC network, with the aim of maximizing the long-term system utility that captures system throughput and user fairness. 
	Then, the proposed algorithm is extended to cases under partial knowledge on device states. The contributions of this paper can be summarized as follows:
\begin{itemize}
	\item 
	We consider a novel and practical problem where offloading decisions and resource allocation are jointly optimized in the NOMA-aided MEC network with the objective of the long-term system utility maximum. Then the problem is extended to a more practical scenario where only partial device states knowledge is obtained at the BS when the BS assists devices in power control.
	
	\item 
	We propose a low-complexity online task offloading algorithm, which firstly decomposes the temporal coupling stochastic problem into a series of per-slot deterministic programming. By introducing a set of auxiliary variables, the non-convex optimization subproblem is reformulated as the equivalent convex form. Accordingly, the large-scale non-convex optimization can be solved with the computation complexity of $\mathcal{O} ( M^{3/2})$.  

	\item 
	We prove that our proposed approach can achieve an $[\mathcal{O}(1/V),\mathcal{O}(V)] $ tradeoff between system utility and stability. The performance loss asymptotically diminishes with the increase of parameter $V$, and the asymptotic optimality of the proposed algorithm is preserved under partial device knowledge.  

	\item
	Corroborated by extensive simulations, the proposed algorithm can substantially elevate system utility compared with benchmarks. Specifically, it outperforms the OFDMA-based scheduling and static strategy up to 85\% and 25\%,  respectively, and reduce the total signaling overhead by 50\%. The asymptotic optimality is validated under different network control parameters and partial device knowledge. 
\end{itemize}

The rest of this paper is organized as follows. In Section \uppercase\expandafter{\romannumeral2}, the system model of NOMA-aided MEC networks is presented. In Section \uppercase\expandafter{\romannumeral3}, we formulate an optimization problem for long-term system utility maximization. Then an online scheduling algorithm is proposed in Section \uppercase\expandafter{\romannumeral4}, and the optimality analysis is given in Section \uppercase\expandafter{\romannumeral5}. Finally, Section \Rmnum{6} provides the simulation results, which is followed by conclusion in Section \Rmnum{7}.

\section{System Model}
Consider a NOMA-aided MEC system wherein a BS (co-located with an edge server) serves $M$ mobile devices, as shown in Fig. 1. The system operates in a slotted structure, i.e., $t \in \mathcal{T}=\{0,1,\cdots\}$, and the slot length is $\Delta T$. The computation-intensive tasks generated by devices can either be processed locally or offloaded to the BS and processed by MEC server through NOMA. Data processing on the device takes place in two modules, i.e., wireless module and computation module. Wireless module maintains an offloading queue for storing the tasks that will be offloaded. Computation module maintains a local computing queue for storing data that will be processed locally. At the BS side, superimposed signals from multiple devices are first decoded in the SIC decoding module, and then executed in the data processing module. Besides, device state information such as task arrivals, queue status and channel state information, are collected by the BS and used to make resource scheduling strategy for devices. Further, the processed results and scheduling instructions are sent to devices in the wireless module. For ease of reference, notations in this paper are listed in Table II.

\begin{table}[t]   
	\caption{Summary of Key Notations}
	\centering   
	\begin{tabular}{ll}   
		\hline 
		\textbf{Notation} & \textbf{Description} \\   
		\hline 
		$M$ & The number of devices \\ 
		$g_{m,t}$ & Channel gain from device $m$ to BS \\ 
		$A_m(t)$ & The size of task arriving at device $m$ \\ 
		$C_m(t)$ & CPU cycles for processing one bit of data \\ 
		$\rho_{m,t}$ & Offloading decision of device $m$ \\ 
		$I_m(t)$ & Interference of device $m$ \\ 
		$p_m(t)$ & Uplink transmit power of device $m$ \\ 
		$p_{m,\text{loc}}(t)$ & Power of local computing on device $m$ \\ 
		$p_{m,\text{tot}}(t)$ & Total power for processing task on device $m$ \\ 
		$W$ & Channel bandwidth \\ 
		$n_0(t)$ & Additive white Gaussian noise power \\ 
		$R_m(t)$ & Achievable uplink transmission rate of device $m$  \\ 
		$f_{m}(t)$ & CPU frequency on device $m$ \\ 
		$f_{m,\text{max}}(t)$ & Maximum CPU frequency on device $m$ \\ 
		$\kappa_m$ & Power-scaling parameter \\ 
		$Q_{m,\text{loc}}(t)$ & Local computing queue on devices $m$\\ 
		$Q_{m,\text{off}}(t)$ & Task offloading queue on device $m$ \\ 
		$Q_{m,p}(t)$ & Virtual power queue on device $m$\\ 
		$Q_{\text{BS}}(t)$ & Task processing queue at BS \\ 
		$w_m$ & Weighing parameter of device $m$ \\ 
		$U_m(t)$ & Utility of device $m$  \\ 
		$p_{\text{ave}}$ & The expected average power budget \\ 
		$p_{\text{max}}$ & Maximum transmit power  \\ 
		$S_m(t)$ & State information of device $m$ \\ 
		$\widehat{S}_m(t)$ & Approximate state information of device $m$\\ 
		\hline                    
	\end{tabular}
\end{table}

\subsection{Communication Model}
Let ${\mathcal{M}} = \left\{ {1,2, \cdots ,M} \right\}$ denote the set of $M$ devices. The wireless channels between devices and the BS are assumed to follow independent and identically  distributed (i.i.d.) block fading \cite{Wu2019Online}. Specifically, the channel gains vary between different time slots, while staying unchanged within each offloading period. Denote by $g_{m,t}$ the channel gain from device $m$ to BS at time slot $t$. Without loss of generality, for the $t-$th time slot, we assume the channel gains are sorted in a non-increasing order, i.e., ${g_{1,t}} \ge {g_{2,t}} \ge  \cdots  \ge {g_{M,t}}$. 

Let ${\Gamma}_{m,t} = \left\lbrace  A_m(t), C_m(t) \right\rbrace  $ represent the computation task generated by device $m$ at time slot $t$. $A_m(t) \le A_{m,\max}$ is the size of tasks (in bps) arriving at device $m$, and $C_m\left(t \right) $ (cycles/bit) is the CPU cycles required to process a bit of data for the task  during slot $t$. We assume that ${\Gamma}_{m,t}$ is generated in the i.i.d. manner, which can capture the stochastic and intermittent nature of the task arrivals. $\rho_{m,t} \in \left\lbrace 0,1 \right\rbrace $ is the offloading decision indicator for device $m$. $\rho_{m,t} = 0$ if task $\Gamma_{m,t}$ is executed locally, and $\rho_{m,t} = 1$ otherwise.

Exploiting NOMA technique, multiple devices can simultaneously transmit data to the BS using the same PRB, and therefore the superimposed signal received at the BS contains not only the desired signal, but also interferences from co-sharing devices. With the SIC technique, the signals are iteratively decoded in the order of transmitter channel gains \cite{Zhai2019Delay,Dai2015Non}. In each iteration, the BS decodes the signal with the strongest channel gain of the remaining devices, and removes it from the superimposed signals. When decoding the signal of device $m$, the interference of the weaker signals from devices $m+1$ to $M$, denoted by $I_m(t)$, can be written as 
\begin{align}
& {I_m}(t) = \sum\nolimits_{i = m + 1}^M {{g_{i,t}}{p_i}(t)}, \notag \\
& 0 \le {p_m}(t) \le {p_\text{max}},\forall m \in {\mathcal{M}}, \label{Constraint-3}
\end{align}
where $p_m(t)$ is the uplink transmit power of device $m$ and bounded by  ${p_\text{max}}$.

The available uplink NOMA transmission rate of device $m$ at time slot $t$ is given by \cite{Shannon1938Asymbolic}
\begin{equation}
\label{uplink_rate}
{R_m}(t) = W{\log _2}\left( {1 + \frac{{g_{m,t} {p_m}(t)}}{ {I_m}(t) + n_0(t) }} \right),
\end{equation} 
where $W$ is the channel bandwidth and $n_0 (t) $ is the i.i.d. additive white Gaussian noise (AWGN) power. 

\subsection{Queuing Model}
 
 \begin{figure}[t] 	\label{scenario}
 	\centering
 	\includegraphics[width=3.5in]{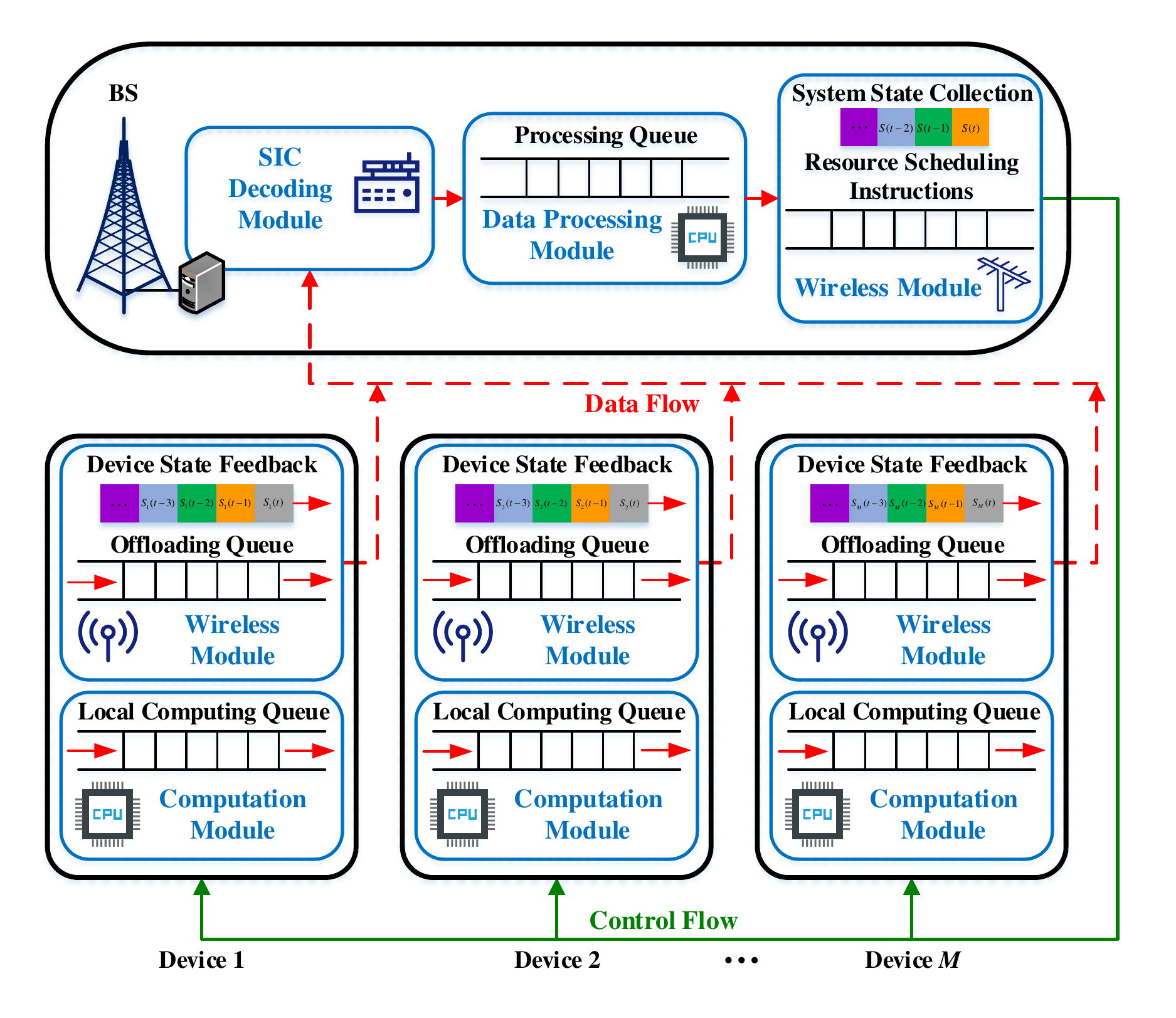} 
 	\caption{The NOMA-aided MEC offloading system.}
 \end{figure}

	The NOMA-aided MEC system is modeled as a cascaded queuing model as shown in Fig. 1. In particular, each device maintains a local computing queue $Q_{m,\text{loc}}(t) $ for unprocessed local workloads and a task offloading queue $Q_{m,\text{off}}(t) $ for the workloads waiting to be offloaded, while the BS maintains a task processing queue $Q_{\text{BS}}(t) $ for buffering offloaded workloads from devices. The above queues are the data structures holding messages in a first-in, first-out (FIFO) fashion \cite{Kua2017Using}.
 	
	The input to  $Q_{m,\text{loc}}(t) $ at each time slot is the portion of workloads for local computing, i.e., $\left(  1 - \rho _{m,t}\right) {A_m}(t)$, while the output is the amount of data processed by local CPU, denoted as ${\mu _m}(t)$. Therefore $Q_{m,\text{loc}}(t) $ is updated by:
	\begin{equation}
	\label{localcomputing_queue}
	{Q_{m,\text{loc}}}(t+1) = {\left[ {{Q_{m,\text{loc}}}(t) - {\mu _m}(t)} \right]^ + } + \left( 1 - \rho_{m,t} \right) {A_m}(t),
	\end{equation} 
	where ${\left[ x \right]^ + }$ denotes $\max \left\{ {x ,0} \right\}$.
 
	By adjusting the transmit power $p_m(t) $, device $m$ can offload different amounts of data to the BS. Constrained by the limited channel capacity, at most ${R_m}(t)$ amounts of data from device $m$ can be transmitted to the BS during a time slot interval. Therefore, $Q_{m,\text{off}}(t) $ dynamics are given by 
	\begin{align}
	\label{offloading_queue}
	{Q_{m,\text{off}}}(t+1) = {\left[ {{Q_{m,\text{off}}}(t) - {R_m}(t)} \right]^ + } +  {\rho _{m,t}}{A_m}(t).
	\end{align} 
 
 	The arrivals on  $Q_{\text{BS}}(t) $ at the BS consist of all the offloaded workloads from different devices within a time slot, i.e., $\sum\nolimits_{m = 1}^M {{R_m}(t)}$. The output of $Q_{\text{BS}}(t)$ is the size of data processed at the BS, denoted by
	 $\mu _{\text{BS}}$. Consequently, $Q_{\text{BS}}(t) $ is updated by 
	\begin{align}
	\label{BScomputing_queue}
	{Q_{\text{BS}}}(t+1) = {\left[ {{Q_{\text{BS}}}(t) - {\mu _{\text{BS}}}(t)} \right]^ + } + \sum\nolimits_{m = 1}^M {{R_m}(t)}.
	\end{align}

\subsection{Computation Model}
Each device chooses either to process tasks locally or offload for remote execution. Regarding local computing, we denote ${f_m}(t)$ as the CPU frequency on device $m$ satisfying
\begin{align}
\label{Constraint-2}
0 \le {f_m}(t) \le {f_{m,\text{max }}},\forall m \in {\mathcal{M}},
\end{align}
where $f_{m,\text{max }}$ is the maximum CPU frequency on device $m$.

At the BS side, $f_{\text{BS}}(t)$ is allocated for processing offloaded tasks from devices. Therefore, the sizes of tasks that can be computed locally at device $m$ and remotely at the BS during slot $t$ can be given by  
\begin{align}
& {\mu _m}(t) = \frac{{{f_m}(t)\Delta t }}{{{C_m (t) }}},  \label{local_service_rate} \\
& {\mu _{\text{BS}}}(t) =  \sum\nolimits_{m = 1}^M{\frac{{  { {f_{\text{BS}}}(t)\Delta t }}}{{{C_m}(t)}}} \label{BS_service_rate}.
\end{align}

Using dynamic voltage frequency scaling \cite{Vogeleer2014The}, the power consumption of local processing on device $m$ is
\begin{equation}
\label{local_power}
{p_{m,\text{loc}}}(t) = {\kappa _m}f_m^{3}(t),
\end{equation}
where ${\kappa _m}$ is the power-scaling parameter depending on the CPU architecture of the device \cite{Hua2019Energy}.

	Following the FIFO policy, if a task is decided to be computed locally, it will be placed at the bottom of $Q_{m,\text{loc}}(t)$, otherwise, it will be placed at the bottom of $Q_{m,\text{off}}(t)$. Local computing and data offloading are performed simultaneously at devices. Tasks being processed at current time slot come from the top of queues. Therefore, if $|Q_{m,\text{loc}}(t)| \neq 0$ and $|Q_{m,\text{off}}(t)| \neq 0$, the total power consumption consists two parts, i.e., local computing power $p_{m,\text{loc}}(t)$ and offloading transmission power $p_m(t)$. On the other hand, if either of the queues in $Q_{m,\text{loc}}(t)$ and $Q_{m,\text{off}}(t)$ is empty, the corresponding power is not required. To illustrate this, we introduce the queue state indicators $x_{m,\text{loc}}(t), x_{m,\text{off}}(t) \in \{0,1\}$, and define $x_{m,\text{loc}}(t)=1$ if $|Q_{m,\text{loc}}(t)| \neq 0$, and $x_{m,\text{loc}}(t)=0$, otherwise. $x_{m,\text{off}}(t)$ is defined in the same way for $Q_{m,\text{off}}(t)$. Besides, the computation results from the BS are usually of small size \cite{You2017Energy,Wang2018Joint,Nouri2018NOMA}, and the standby power accounts for a relatively small proportion of overall power consumption by standby power management techniques such as autonomous normally-off power transition\cite{Hayashikoshi2017Low}, adaptive data retention voltage regulating scheme\cite{Huang2018Ultra}, and power-gating and source bias design technique\cite{Nii2019Ultra}. Therefore, by neglecting the power consumption of downlink reception and standby mode on devices, the total power consumption can be given by
	\begin{align}
	\label{total_power}
	p_{m,\text{tot}}(t) = x_{m,\text{loc}}(t) p_{m,\text{loc}}(t) + x_{m,\text{off}}(t) p_m(t).
	\end{align}
	
	Considering the worst case, where both local computing queue and offloading queue are not empty, the total power consumption $p_{m,\text{tot}}(t)$ at time slot $t$ can be reduced to $p_{m,\text{tot}}(t) = p_{m,\text{loc}}(t) + p_m(t)$.

\bigskip
\section{Problem Formulation}

\subsection{System Utility And Queue Stability}
	Throughput is a fundamental indicator of the quality of service in the IoT, and it can characterize the feasible data dissemination rate of the network. By considering the fairness among devices, $U_m(t)$ is defined as the utility function of device $m$ at time slot $t$ as follows:
	\begin{align}
	\label{utility_function}
	{U_m}\left( {{t}} \right) = {w_m}\log \left( {1 + x_{m,\text{off}}(t) {R_m(t) }} \right), \forall m \in \mathcal{M},
	\end{align} 
	where $w_m$ is a weighting parameter of device $m$. Multiplying $R_m(t)$ by $x_{m,\text{off}}(t)$ implies that there is no need to transmit any workloads if the offloading queue is empty.

	$U_m(t)$ is monotonically increasing and concave w.r.t. $R_m$, since it is a logarithmic function of $R_m$. The reason for formulating utility function in this form is that it can balance the network throughput and fairness by discouraging individual device to occupy excessive resource for little utility \cite{Niyato2011Optimal,Truong2014AStochastic,Neely2010Stochastic,Lyu2017Optimal}. 

Furthermore, the system utility is defined as the sum of all individual utility $U_m \left( t\right) $, which is expressed as follows
\begin{align}
\label{total_utility}
U (t)  = \sum\nolimits_{m{\rm{ = }}1}^M {{U_m (t) }} .
\end{align}

	From (\ref{utility_function}), a positive $x_{m,\text{off}}(t)$ and a larger $R_m(t)$ contribute to a greater utility. On one hand, from the device's perspective, each device is eager to acquire the computation resource of the BS. However, this does not mean that the more data to be offloaded, the greater system utility is. If we are greedy to transmit offloaded data for aggressive utility improvement, both $Q_{m,\text{off}}(t)$ and $Q_{\text{BS}}(t)$ will increase unboundedly. Consequently, it can result in excessive delay and poor user experience. Therefore, there is a tradeoff between the offloading and local computing. As illustrated in Fig. 1, the NOMA-aided MEC network is modeled as a queue system. According to \emph{Little's Law} \cite{Babu2003Introduction}, the average task processing delay is proportional to the queue lengths and follows a power law against the task arrivals at both devices and BS. Therefore, queue stability is a requirement to implement limited service latency. On the other hand, given $x_{m,\text{off}}(t)=1$, a higher transmit power results in greater transmission rate. However, if users increase their transmit power myopically, the inter-user interference will also enlarge, which hinders the transmission rate and causes enormous waste of energy. Thus, the power constraint should be taken into consideration when optimizing the system utility.  

Due to the stochastic task arrivals, dynamic queue backlogs, as well as the variation of the wireless channel conditions in the considered system, we focus on the long-term system performance rather than the instantaneous performance. To illustrate these, we give the following  definitions.

\begin{definition}
	The time average expectation of a discrete-time stochastic processes $X(t) $ is defined as 
	\begin{equation}
	\label{time_average_expectation}
	\overline {X(t)} = \mathop {\lim }\limits_{T \to \infty } \frac{1}{T}\sum\nolimits_{t = 0}^{T-1} {\mathbb{E}\left[ {X(t)} \right]}, 
	\end{equation}
	where $\mathbb {E}\left[  X(t)  \right] $ is the expectation of $X(t) $.
\end{definition}

\begin{definition}
	A discrete-time process $Q(t) $ is mean rate stable if
	\begin{equation}
	\label{mean_rate_stable}
	\mathop {\lim }\limits_{t \to \infty } \frac{1}{t}\mathbb{E}\left[ | {Q}(t)  |  \right]  =0.
	\end{equation}
\end{definition}

	The stability of a network is equivalent to the stability of all individual queues, such that all the arrived workloads can be completed within the finite delay \cite{Li2019Lyapunov,Zhai2018Energy}. Accordingly, we have the following queue stability constraints:
	\begin{align} 
	&\mathop {\lim }\limits_{t \to \infty } 
	\frac{1}{t}{\mathbb{E} [  | {Q_{m,\{\text{loc},\text{off}\}}}(t)   |  ]}   =0, \ \forall m, \label{stability_Q_m_off}  \\
	&\mathop {\lim }\limits_{t \to \infty } \frac{1}{t}  {\mathbb{E} [ {  | {Q_{\text{BS}}}(t)   |  }  ]}  = 0. \label{stability_Q_BS} 
	\end{align}

\subsection{System Utility Maximum Problem}

With the aim of maximizing the long-term average system utility by jointly optimizing the offloading decision, CPU frequency and power control on devices, the stochastic optimization problem can be formulated as
\begin{subequations} \label{problem}
	\begin{eqnarray}
	&\underset{\boldsymbol{x}_t}{\max}& \sum \nolimits_{m \in \mathcal{M}} {\overline{U_m(t)}}\\
	&s.t.& \rho _{m,t} \in \{0, 1\} , \ \forall m,t, \label{rho_constraint}\\
	&{}& \overline{{p}_{m,\text{tot}}} \le {p_{\text{ave}}}, \ \forall m,t, \label{transmit_power}\\
	&{}&  (\ref{Constraint-3}), (\ref{Constraint-2}), (\ref{stability_Q_m_off}) \text{~and~} (\ref{stability_Q_BS}), \label{queue_stable}
	\end{eqnarray}
\end{subequations}
where
$\boldsymbol{x}_t = \{ \boldsymbol{\rho}_t, \boldsymbol{f} (t), \boldsymbol{p} (t) \}$ is the offloading strategy for the considered system in each time slot $t$, ${\boldsymbol{\rho }_t} = \{ \rho _{1,t}, \cdots , \rho _{M,t} \}$, $\boldsymbol{f} (t) = \{ {{f_1} (t), \cdots ,{f_M}} (t) \} $ and $\boldsymbol{p}(t) = \{ {{p_1}(t), \cdots ,{p_M}(t) } \} $. Constraint (\ref{rho_constraint}) is the offloading decision constraint. Constraint (\ref{transmit_power}) implies that the time average overall power consumption on each device is within the expected average power budget $p_{\text{ave}}$. Constraints (\ref{Constraint-3}) and (\ref{Constraint-2}) guarantee that the allocated transmit power and instantaneous CPU frequency cannot exceed their maximums. Constraints (\ref{stability_Q_m_off}) and (\ref{stability_Q_BS}) are the guarantee of queue stability of the system.

The objective of problem (\ref{problem}) is a long-term stochastic optimization where the multi-dimensional decision variables are adaptive to the dynamic network. An offline optimization of problem (\ref{problem}) requires comprehensive system knowledge, which is impossible in practice due to the highly stochastic and unpredictable characteristics of the considered system. Keeping the above challenges in mind, we are motivated to propose an online task offloading algorithm to jointly optimize $\boldsymbol{\rho}_t, \boldsymbol{f} (t), \boldsymbol{p} (t)$ with excellent scalability and low computation complexity. 

\section{Online Optimization for Task Offloading}

In this section, we develop an online task offloading algorithm based on Lyapunov optimization whereby the original problem (\ref{problem}) is decoupled into a series of deterministic per-time slot programming. Further, the application of our proposed algorithm is extended to a more practical scenario by considering partial device knowledge.

\subsection{The Formulation of Lyapunov Optimization}
	Considering the time averaged power consumption constraint (17c), if inequation $\overline{{p}_{m,\text{loc}}} + \overline{{p}_{m}} \le p_{\text{ave}}$ holds, constraint (17c) is valid for all cases of $Q_{m,\text{loc}}(t)$ and $Q_{m,\text{off}}(t)$. Therefore, to satisfy constraint (17c), we introduce a virtual queue $Q_{m,p}(t)$ to transform it into the following equivalent form:
	\begin{align}
	\label{virtual_queue_of_power}
	{Q_{m,p}}(t+1) = {\left[ {{Q_{m,p}}(t) - {p_{m,\text{ave}}}} \right]^ + } + {p_{m,\text{tot}}}(t)	,\forall m \in \mathcal{M}.
	\end{align}
\begin{lemma}
	\label{mean_rate_satble}
	Constraint (\ref{transmit_power}) is satisfied, if $Q_{m,p}(t) $ is mean rate stable.
\end{lemma}

\begin{IEEEproof}
Please refer to Appendix A.
\end{IEEEproof}

After introducing the virtual queue, we define the Lyapunov function as
\begin{equation} \label{L}
\begin{aligned}
L( \boldsymbol{\Theta} (t) ) \buildrel \Delta \over= \frac{1}{2} \sum \nolimits_{m = 1}^M & [ Q_{m,\text{off}}^2 (t) + Q_{m,\text{loc}}^2 (t)  \\ 
& + Q_{m,p}^2(t) ] + \frac{1}{2} Q_{\text{BS}}^2 (t),
\end{aligned}
\end{equation}
where $\boldsymbol{\Theta} (t)  = \{ {Q}_{1,\text{off}}(t) ,\cdots, {Q}_{M,\text{off}}(t)  ,\cdots,{Q}_{M,\text{loc}}(t),\cdots,$ ${Q}_{M,p}(t) , {Q}_{\text{BS}}(t) \} $ is a concatenated vector of all actual and virtual queues in the system. From equation (\ref{L}), the conditional Lyapunov drift is given as 
\begin{align}
& \Delta L(t) \buildrel \Delta \over = \mathbb{E} \left[ {\left. {L\left( {\boldsymbol{\Theta}(t+1)} \right) - L\left( {\boldsymbol{\Theta}(t)} \right)} \right|\boldsymbol{\Theta}(t)} \right]. \label{L_drift}
\end{align}

To solve problem (\ref{problem}), we map the objective $U(t)$ as a penalty, and introuduce the Lyapunov drift-plus-penalty function:
\begin{align}
\label{drift_plus_penalty}
{\Delta _V}\left( {\boldsymbol{\Theta}(t)} \right) \buildrel \Delta \over = \Delta L(t) - V\mathbb{E}\left[ {\left. {U(t)} \right|\boldsymbol{\Theta}(t)} \right],
\end{align}
where $V$ is a non-negative constant control parameter set by the NOMA-aided MEC system to balance the system utility and queue stability. The following Lemma \ref{dpp_up_bounded} gives the upper bound of ${\Delta _V}\left( {\boldsymbol{\Theta}(t)} \right)$.	

\begin{lemma}
	\label{dpp_up_bounded}
	For any offloading schedule and for any time, given any $\boldsymbol{\Theta}(t)$, (\ref{drift_plus_penalty}) is upper bounded by 
	\begin{align} \label{up_bound}
	{\Delta _V} ({\boldsymbol{\Theta} (t ) } ) & \le \sum_{m = 1}^M 
	\left\{ {} \right.      
	{{Q_{m,\text{off}}} (t )  \mathbb{E} \left[ {\left. {{\rho _{m,t}}{A_m} (t )  - {R_m} (t ) } \right|\boldsymbol{\Theta} (t ) } \right]} \notag \\
	& + {Q_{m,\text{loc}}} (t )  \mathbb{E} [ {\left. { ({1 - {\rho _{m,t}}} ) {A_m} (t )  - {\mu _m} (t ) } \right| \boldsymbol{\Theta}  (t ) } 
	\notag \\ 
	& + {Q_{m,p}} (t )   \mathbb{E} \left[ {\left. {{p_{m,\text{tot}}} (t )  - {p_{\text{ave}}}} \right| \boldsymbol{\Theta}  (t ) } \right] \left. {} \right\} \notag \\
	& + {Q_{\text{BS}}} (t )   \mathbb{E} [  \sum\limits_{m = 1}^M {{R_m} (t) }  - {\mu _{\text{BS}}} (t)  |  \boldsymbol{\Theta}  (t )  ] \notag \\
	&  + B - V\mathbb{E}\left[ {\left. {U (t ) }  \right| \boldsymbol{\Theta} (t ) } \right],
	\end{align}
	where
	$B \buildrel \Delta \over = \frac{1}{2} \sum \nolimits_{m=1}^{M} { ( {2A_{\max }^2 + 2R_{m,\max }^2 + \mu _{m,\max }^2 + 2 p_\text{max}^2} ) }  + \frac{1}{2} \mu_{\text{{BS}},\max}^2$.
\end{lemma}

\begin{IEEEproof}
	Please refer to Appendix B.
\end{IEEEproof} 

Exploiting Lemma \ref{dpp_up_bounded}, we are motivated to minimize the right-hand-side of  (\ref{up_bound}), and reformulate the following deterministic per-slot problem subject to the instantaneous constraints.
\begin{subequations} \label{problem_formulation}
	\begin{eqnarray}
	&\mathop {\min }\limits_{\boldsymbol{x}_t}& {H_1 (\boldsymbol{\rho}_{t} ) + H_2 ( {\boldsymbol{f}} (t) ) + H_3 ( \boldsymbol{p}} (t) )  \\
	&s.t.&  (\ref{Constraint-3}), (\ref{Constraint-2}) \text{~and~} (\ref{rho_constraint}),
	\end{eqnarray}
\end{subequations}
where
\begin{subequations}
	\begin{align}
	{H_1}\left( {\boldsymbol{\rho }}_t \right) \ \ \ & =  \sum\limits_{m = 1}^M { \{ \left[ {{Q_{m,\text{off}}}(t) - {Q_{m,\text{loc}}}(t)} \right]{A_m}(t){\rho _{m,t}}}   \notag\\
	&  + Q_{m,\text{loc}}(t) A_m(t) \}, \label{H1}  \\
	{H_2}\left( {\boldsymbol{f}(t)} \right) &=    \sum\limits_{m = 1}^M \{ {Q_{m,p}}(t){\kappa _m}f_m^3(t)  -  Q_{m,p}(t) p_{\text{ave}} \notag\\
	&  - {\varepsilon Q_{m,\text{loc}}(t) {f_m}(t) }/{C_m}(t) \} , \label{H2} \\
    {H_3}\left( {\boldsymbol{p}(t)} \right) & =  \sum\limits_{m = 1}^M \{ {Q_{m,p}}(t){p_{m}}(t) \notag\\
	& - V{w_m} \log \left( {1 + x_{m,\text{off}}(t){R_m}(t)} \right)  \notag\\
	& +  \left[ {{Q_{\text{BS}}}(t) - {Q_{m,\text{off}}}(t)} \right]{R_m}(t) \}. \label{H3} 
	\end{align}
\end{subequations}

Note that $\boldsymbol{\rho}_{t},\boldsymbol{f}(t), \boldsymbol{p} (t)$ are decoupled in both the objective and constraints in  (\ref{problem_formulation}), and therefore can be addressed by dealing with the following three subproblems separately.

\subsection{Online Offloading Scheduling}

\subsubsection{Optimal Offloading Decision $\boldsymbol{\rho}_t$} 

The optimal solution of offloading decision $\boldsymbol{\rho_t}$ can be obtained by solving
\begin{subequations}\label{offloading}
	\begin{align} 
	&\mathop {\min } \limits_{ \boldsymbol {\rho}_t  }  \quad {H_1}\left( {\boldsymbol{\rho _t}} \right) \\
	&\mathop {s.t.} \quad  {\rho _{m,t}} \in \left\{ {0,1} \right\} ,\forall m, \label{rho_constraint_in_H1}
	\end{align}
\end{subequations}
where $\boldsymbol {\rho}_t$ is a binary variable that makes (\ref{offloading}) a zero-one integer programming. By employing a relaxation approach to relax $\boldsymbol {\rho}_t$ into $\left[ 0,1 \right] $, problem (\ref{offloading}) can be simplified to a sequential linear programming. By utilizing the monotonicity of ${H_1}( {\boldsymbol{\rho}_t} )$, the following theorem claims that there is no optimality loss after constraint relaxation.
\begin{theorem} 
	\label{rho_without_loss}
	The optimal offloading decision $\boldsymbol{\rho}_{t}^*$ is given as follows without optimality loss in (\ref{offloading}).
	\begin{align}
	\label{optimal_offloading_decision}
	\rho_{m,t}^* = \left\{ \begin{array}{l}
	{0}, \quad {  Q_{m,\text{off}}}(t) \ge {  Q_{m,\text{loc}}} (t),\\
	1, \quad {\rm otherwise}.
	\end{array} \right.
	\end{align}
\end{theorem}

\begin{IEEEproof}
With the knowledge of queue backlogs and task arrivals in the $t$-th time slot, ${H_1}\left( {\boldsymbol{\rho }_t} \right) $ is a linear function of continuous variables $ { {\rho _{m,t}}} \in \left[ 0,1\right]  $, where the minimum is always met at the boundary of the feasible region, i.e., $\rho _{m,t} = 0$ if $Q_{m,\text{off}}(t) \ge {  Q_{m,\text{loc}}} (t)$, and vice versa. Therefore, after relaxation,  ${\boldsymbol{\rho }_t^*}$ is optimal without violating  (\ref{rho_constraint_in_H1}).
\end{IEEEproof}

\textbf{Theorem \ref{rho_without_loss}} indicates that each device endeavors to pursue queue stability via appropriate offloading decisions. In other words, if the backlog size of $Q_{m,\text{off}}(t)$ is larger than $Q_{m,\text{loc}}(t)$, the device tends to local computing to prevent network congestion and vice versa. 

\subsubsection{Optimal CPU Frequency $\boldsymbol{f}(t) $ }
The optimal CPU frequency on devices in the $t$-th time slot can be achieved by addressing the following subproblem:
\begin{subequations} \label{frequency}
	\begin{eqnarray}
	&\mathop {\min }\limits_{{\boldsymbol{f} (t) }} & {H_2}\left( {\boldsymbol{f}(t)} \right) \\
	&\mathop  {s.t.}& 0 \le {f_m}(t) \le {f_{m,\text{max} }} ,\forall m.
	\end{eqnarray}
\end{subequations}

Since the objective in (\ref{frequency}) is a polynomial function of $\boldsymbol{ f }(t)$ on a continuous definition domain, the optimal $\boldsymbol{f} ^*(t)$ is achieved at either the boundary points or the stationary points, which is given by
\begin{align}
\label{optimal_local_CPU}
f_m^*(t) = \min \left\{ {f_{m,\text{max} }, \ \sqrt {\frac{{{  Q_{m,\text{loc}}(t) }\varepsilon }}{{3{  Q_{m,p}(t) }{C_m \left(t \right) }{\kappa _m}}}} } \ \right\}.
\end{align}

When $f_m (t) \le f_{m,\text{max}}$, the optimal CPU frequency allocation is proportional to the square root of $Q_{m,\text{loc}}(t)$ and is the inverse of $Q_{m,p}(t)$. Such a result can be explained as follows. With the increase of $Q_{m,\text{loc}}(t)$, 
more CPU resources are demanded to prevent excessive queue backlogs. In contrast, a larger $Q_{m,p}(t)$ means that the power budget on the device is sufficient enough, which inspires the device to offload for utility maximum. 

From (\ref{optimal_offloading_decision}) and (\ref{optimal_local_CPU}), it is noteworthy that devices choose optimal control actions $\boldsymbol{\rho }_t^*$ and $\boldsymbol{f}^*(t)$ by observing their respective queue state in each time slot $t$ independently without the assistance of BS. 

\subsubsection{ Optimal Transmit Power $\boldsymbol{p} (t)$}

At each time slot, devices allocate transmit power $\boldsymbol{p} (t)$ by solving the following subproblem:
\begin{subequations} \label{power_allocation}
	\begin{eqnarray}
	&\mathop {\min }\limits_ {  \boldsymbol{p}(t) } & {H_3}\left( {\boldsymbol{p}(t)} \right) \\	
	&{s.t.} &  0 \le  {p_{m} }(t) \le {p_\text{max}} ,\forall m.
	\end{eqnarray}
\end{subequations}

In order to obtain the optimal power control, the real-time states of overall devices should be coordinated comprehensively at the BS. However, real-time feedback from devices to the BS contends for communication resources with offloading, which impairs system performance, especially in large-scale networks. 

In a practical scenario where any device that has fed back at time slot $t$ should idle $n\Delta t$ periodic time for the next feedback, where $n \in \{0, \cdots , T\}$ is a constant obtained by historical experience so as to harmonize system overhead with time validity, 
and $T$ is the maximum feedback interval. $S_m (t) = \{ Q_{m,\text{loc}}(t), Q_{m,\text{off}}(t), Q_{m,p} (t) \}$ is referred to as the state information of device $m$ at time slot $t$. Accordingly, the approximate state information is expressed as 
\begin{equation}
\label{approximate}
{\widehat S_m}(t) = {S_m}(t-n\Delta t).
\end{equation}
 
\begin{algorithm}[t]
	\caption{Online Optimization for Task Offloading}
	\label{algorithm_1}
	\begin{algorithmic}[1]
		\renewcommand{\algorithmicrequire}{\textbf{For each device $m$:}}
		\renewcommand{\algorithmicensure}{\textbf{At the MEC server:}}
		\REQUIRE
		\STATE Acquire $ Q_{m,\text{loc}}(t) $, $ Q_{m,\text{off}}(t) $ and $ Q_{m,p}(t) $. 
		\STATE Conduct the optimal solutions of offloading decision and CPU frequency based on (\ref{optimal_offloading_decision}) and (\ref{optimal_local_CPU}) respectively. 
		\STATE Update $ Q_{m,\text{loc}}(t) $, $ Q_{m,\text{off}}(t) $ and  $ Q_{m,p}(t) $ according to (\ref{localcomputing_queue}), (\ref{offloading_queue}) and (\ref{virtual_queue_of_power}).
		\STATE Feed back $ Q_{m,\text{loc}}(t) $, $ Q_{m,\text{off}}(t) $ and $ Q_{m,p}(t) $ to the MEC server.
		\ENSURE   
		\STATE Acquire $ Q_{\text{BS}}(t) $.
		\STATE Collect feedbacks ${ {\boldsymbol{S}}}(t)$ from a portion of devices, and reserve approximate state ${\widehat {\boldsymbol{S}}}(t)$ for the rest.
		\STATE Schedule the power allocation by solving (\ref{reformulated}).
	\end{algorithmic}
\end{algorithm}

The BS does not always capture complete system knowledge since not all devices feedback in each time slot. Instead, the BS utilizes ${\widehat {\boldsymbol{S}}}(t)=\{ {\widehat S_1}(t),\cdots,{\widehat S_M}(t)\}$ as partial device knowledge for scheduling. It is accepted that partial knowledge implies complete knowledge under ideal condition $n=0$. After introducing partial knowledge, we proceed to resolve problem (\ref{power_allocation}). The non-convexity of (\ref{power_allocation}) is due to the coupling of $\boldsymbol{p}\left(t \right)$ in the objective. To tackle this challenging issue, we introduce auxiliary variables ${\nu _{m,t}} =  \sum\nolimits_{i = 1}^m {{R_i}(t)} $ and $1/g_{0,t} = 0$, and derive the following lemma. 

\begin{lemma} 
	\label{convex}
Problem (\ref{power_allocation}) can be reformulated as the following equivalent form, 
\begin{subequations} \label{reformulated}
	\begin{eqnarray}
	& \min \limits_{ \boldsymbol{\nu}_t } & {H_3}\left( {\boldsymbol{\nu}_t} \right) \\
	&{{\rm s.t.}} & 0 \le  {p_{m} } ( {\boldsymbol{\nu}_t} ) \le {p_\text{max}} ,\forall m,
	\end{eqnarray}
\end{subequations}
where
\begin{subequations}
	\begin{align}
	{H_3}\left( {\boldsymbol{\nu}_t} \right) = &
	\sum\limits_{m = 1}^M {} \{ - {\widehat Q_{m,\text{off}}}(t){R_m}\left( {\boldsymbol{\nu}_t}\right) + {\widehat Q_{m,p}}(t){p_m}\left({\boldsymbol{\nu}_t}\right) \notag \\
	-& V{w_m}\log \left( {1 + {R_m}\left( {\boldsymbol{\nu_t }} \right)} \right)\}  + {\widehat Q_{\text{BS}}}(t){\nu _{M,t}},\\
	{p_m}\left({\boldsymbol{\nu_t }}\right) = & ( {{1}/{{{g_{m,t}}}} - {1}/{{{g_{m-1,t}}}}}  ){2^{{\nu _{m,t}}}} - {1}/{{M{g_{1,t}}}}, \label{p_m_to_nu} \\
	{R_m}\left( {\boldsymbol{\nu_t }} \right) = & {\nu _{m,t}} - {\nu _{m - 1,t}}.  \label{R_m_to_nu}
	\end{align}
\end{subequations}
\end{lemma}

\begin{IEEEproof}
	Please refer to Appendix C.
\end{IEEEproof}

Both (\ref{p_m_to_nu}) and (\ref{R_m_to_nu}) are convex since ${p_m}\left({\boldsymbol{\nu}_t}\right)$ is an exponential function and ${R_m}\left( {\boldsymbol{\nu }_t} \right)$ is linear in ${\boldsymbol{\nu }_t}$, which preserves the convexity when taking the negative logarithm of ${R_m}\left( {\boldsymbol{\nu }_t} \right)$. With the convex objective and constraints, (\ref{reformulated}) is a convex optimization problem, which can be adequately addressed by off-the-shelf solver such as CVX. Based on the above analysis, \textbf{Algorithm 1} summarizes the online optimization for task offloading. 

\subsection{Computational Complexity Analysis}
First, computation for making offloading decisions and CPU frequency allocation in Step $2$ requires a constant time $\mathcal{O}\left(2M \right) $ for each enumerated slot, since the optimal solutions are given in closed forms in (\ref{optimal_offloading_decision}) and (\ref{optimal_local_CPU}). Then, the complexity of a convex algorithm (such as interior-point methods) is $\mathcal{O} (M^{3/2})$ in Step $7$ \cite{Nemirovskii1994Interior}. The polynomial complexity of the proposed algorithm is in marked contrast to the situation for substantial nonconvex optimization problems, for which in the worst case, a number of operations that is exponential in the problem dimensions \cite{Boyd2004Convex}. Since offloading decisions and resource allocation are made in parallel, the proposed algorithm can significantly reduce the complexity in the large-scale NOMA-aided MEC networks, which is given by $\mathcal{O} (M^{3/2})$.  

\section{Analysis of Optimality Loss}

In this section, the optimality loss of our proposed algorithm is analyzed in complete and partial knowledge cases. We denote ${U ^*}$ as the offline optimum of probelm (\ref{problem}), and ${{\widetilde U }^*}$ as the time average optimum based on \textbf{Algorithm 1} under complete system state information. Then we are ready to put forward the following theorem.

\begin{theorem} 
	\label{tradeoff}
	Under complete knowledge of system states at BS, the gap between ${U ^*}$ and ${{\widetilde U }^*}$ is no larger than $B/V$, i.e.,
	\begin{equation}
	{U ^*}  - {{\widetilde U }^*}  \le {B}/{V}.
	\end{equation}
\end{theorem}
The backlogs of all queues in each time slot are strictly upper bounded by
\begin{equation}
\label{consgestion_bound}
\begin{array}{l}
\mathop {\lim }\limits_{T \to \infty } \sup \sum\limits_{t = 0}^{T-1} \sum\limits_{m = 1}^M   \mathbb{E} \{ {{Q_{m,\text{loc}}}(t) + {Q_{m,\text{off}}}(t)} + Q_{m,p}(t) \} \notag \\
\qquad \qquad \qquad \qquad + {Q_{\text{BS}}}(t)   \le  \frac{{ B+ V  (  { {U ^*} -  {{\widetilde U }^*}  }  ) }}{\varepsilon } ,
\end{array}
\end{equation}
where $\varepsilon$ is a positive constant.

\begin{IEEEproof}
	Please refer to Appendix D.
\end{IEEEproof}

\textbf{Theorem \ref{tradeoff}} indicates that by adjusting $V$ to a sufficiently large value, the system utility can be arbitrary close to the optimum achieved by causality-violating offline scheduling. However, the bound on the average queue length grows linearly with $V$. 
Namely, \textbf{Algorithm 1} achieves an $\left[ \mathcal{O}\left(1/V \right) ,\mathcal{O}\left(V \right)  \right] $ tradeoff between system utility and queue backlogs. 

Next, we discuss the optimality loss under partial knowledge. By considering the analytical tractability, the differences between approximate queue lengths at BS and the actual values at devices are analyzed in the worst case, where the largest feedback interval $T$ is encountered. 

\begin{lemma}
	\label{approximate_queue_bound}
	In any time slot, the differences between the approximate and the actual queue backlogs are bounded by
	\begin{align}
	\label{delta_Q}
	| {{\widehat Q}_{\{i\}} ( t  ) - {Q_{\{i\}}} ( t  )} | \le  T {\delta _{\{i\}}}    ,
	\end{align}
	where ${\widehat Q}_{\{i\}}(t)$, $Q_{\{i\}}(t)$ and $\delta_{\{i\}}$ represent all types of approximate and actual queues as well as the maximum difference of the queue length between two adjacent slots, respectively,  
	and specifically, ${\delta _\text{loc}} = \max \{ {A_m}(t),\varepsilon f_{\max }/{C_m}(t) \}, {\delta _\text{off}} = \max \{ {A_m}(t),R_{m,\max} (t) \}$, $\delta _{p} = \max \{ \kappa_m f_{\text{max}}^3 +p_{\text{max}} ,  p_{\text{ave}}\}$ and ${\delta _{\text{BS}}} = \max \{ {\sum\nolimits_{m = 1}^M {{R_{m,\max }}(t)} },{\sum\nolimits_{m = 1}^M{\varepsilon {f_{\text{BS} }}}/{{{C_m}(t)}}}\}$.	
\end{lemma}
\begin{IEEEproof}
	Please refer to Appendix E. 
\end{IEEEproof}

We denote $\widehat{\boldsymbol{x}}_t= \{ \widehat{\boldsymbol{\rho}}_t, \widehat{\boldsymbol{f}} (t), \widehat{\boldsymbol{p}} (t) \}$ as the offloading strategy based on \textbf{Algorithm 1} under partial knowledge  for problem (\ref{problem_formulation}). The following theorem illustrates the upper bound of performance loss issued from partial knowledge.

\begin{theorem} 
	\label{optimality_loss}
	In each time slot, the optimality loss of $\widehat {\boldsymbol{x}}_t$ for (\ref{problem_formulation}) is within a constant $J$, i.e.,
	\begin{equation}
	\sum\nolimits_{i = 1}^3 {{H_i}\left( \widehat {\boldsymbol{x}}_t \right)}  
	- \sum\nolimits_{i = 1}^3 {{H_i} \left(\boldsymbol{x}_t \right)  } \le  J,
	\end{equation}
\end{theorem} 

\begin{IEEEproof}
	Please refer to Appendix F. 
\end{IEEEproof}

From \textbf{\textbf{Theorem \ref{tradeoff}}} and \textbf{Theorem \ref{optimality_loss}}, it is reasonable to deduce that the partial knowledge at the BS conforms to the asymptotic optimality of (\ref{problem}) under our proposed algorithm. The optimality loss is upper bounded per time slot as
\begin{align}
\label{asymptotic_opt}
{{\widehat U }^*}\left( {{  \widehat{{\rm{\boldsymbol x}}}_t}} \right) - {U ^*}  \le \frac{1}{V}\left( {B + J} \right).
\end{align}

Note that the optimality loss arised from partial knowledge increase linearly with the feedback interval $T$, since $J$ is linear to $T$, which coincides with the intuition that longer feedback interval  generates less timely   knowledge at BS, and thus causes a weak system performance. Besides, the proposed algorithm is able to achieve asymptotic optimum in the case where task arrivals, channel states and queue backlogs can be instantaneously exchanged in the NOMA-aided MEC system. 

\begin{table}[t]  
	\caption{Simulation Parameter Settings} \label{simulation_para}
	\centering   
	\begin{tabular}{|l|l|l|}   
		\hline \textbf{Parameter} & \textbf{Description} & \textbf {Value} \\   
		\hline 
		$M$ & The number of devices & $ 500$ \cite{Lyu2017Optimal} \\ 
		\hline 
		$g_{m,t}$ & \tabincell{ll}{Channel gain \\ {from device $m$ to BS}} & $ 10^{-3} d^{-2}$ \cite{Zheng2021Achievable} \\ 
		\hline 
		$A_m(t)$ & \tabincell{ll}{The size of task \\ {arriving at device $m$}} & $U_c[0,10] $kbps \cite{Lyu2017Optimal} \\ 
		\hline 
		$C_m(t)$ & \tabincell{ll}{CPU cycles for processing \\ {one bit of data}} & $6400$ cycles/bit \cite{Zhang2018Energy} \\ 
		\hline 
		$w_m$ &  \tabincell{ll}{Weighing parameter \\ {of device $m$}} & $U_d [0,10]$ \cite{Cao2019Online} \\ 
		\hline 
		$n_0(t)$ & {AWGN power} & $ 10^{-9}$ W  \cite{Wu2019Online} \\ 
		\hline 
		$\kappa_m$ & {Power-scaling parameter} & $10^{-26}$ \cite{Nouri2020Dynamic}\\ 
		\hline 
		$p_{\text{ave}}$ &  \tabincell{ll}{The expected average \\ {power budget}} & $ 250 $ mW \cite{Mao2017Stochastic} \\ 
		\hline 
		$p_{\text{max}}$ & {Maximum transmit power} & $ 500 $ mW \cite{Mao2017Stochastic} \\ 
		\hline 
		$f_{m,\text{max}}$ & \tabincell{ll}{Maximum CPU frequency \\ {on device $m$}} & $4 \times 10^{8}$ cycles/s \cite{Qin2019Power} \\ 
		\hline                  
	\end{tabular} 
\end{table}

\section{Simulation Results} 

\begin{figure}[t] 
	\centering 
	\includegraphics[width=2.5in]{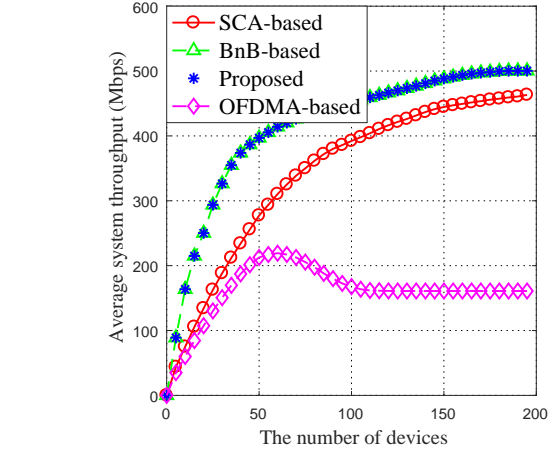} 
	\caption{The average system utility of four offloading scheduling versus the number of devices, with the assumption of the overall device knowledge.} 
	\label{system_utility_vs_arrival}
\end{figure}
\begin{figure}[t] 
	\centering 
	\includegraphics[width=2.5in]{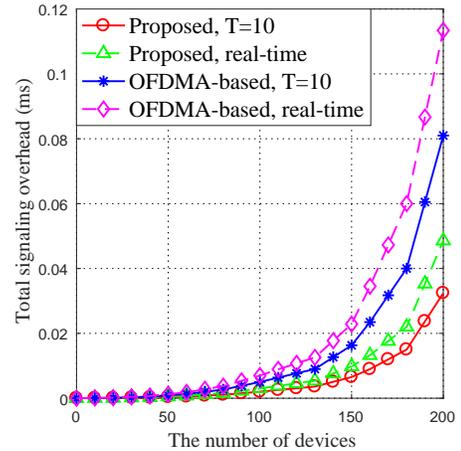} 
	\caption{The signaling overhead as the number of devices increases. The proposed algorithm can achieve the minimum signaling overhead, validating its scalability under large-scale networks.} 
	\label{overhead_vs_device}
\end{figure}

\begin{figure*}[t]
	\centering
	\subfigure[Average system throughput vs. $V$.]{
	\begin{minipage}[t]{0.31 \textwidth}
		\centering
		\includegraphics[width=2.5 in]{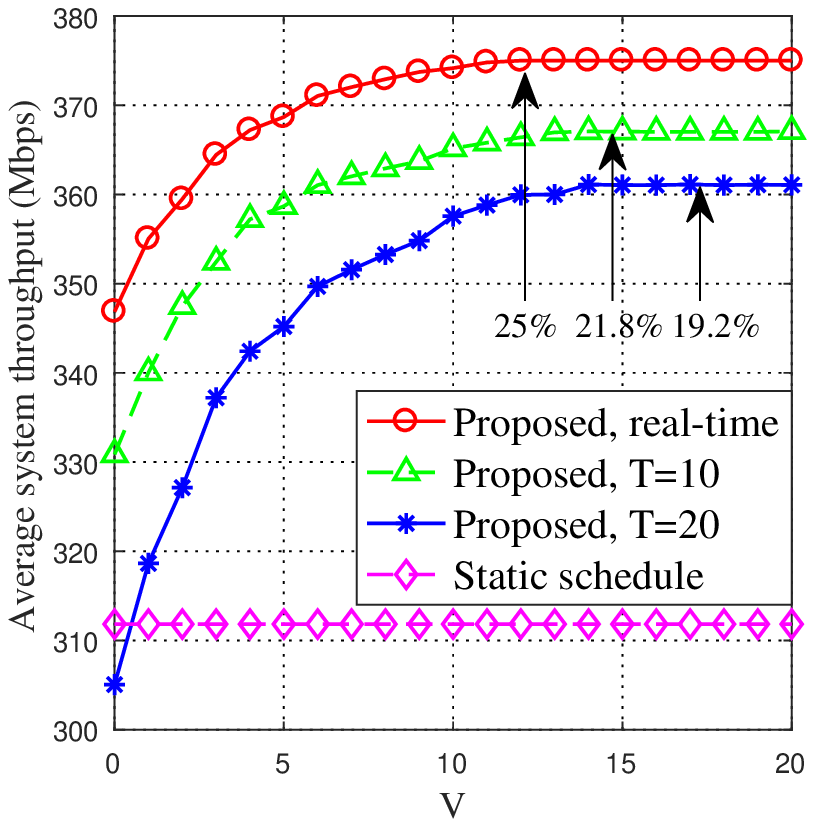}
	\end{minipage}
}
	\subfigure[Total queue lengths vs. $V$.]{
	\begin{minipage}[t]{0.31 \textwidth}
		\centering
		\includegraphics[width=2.5 in]{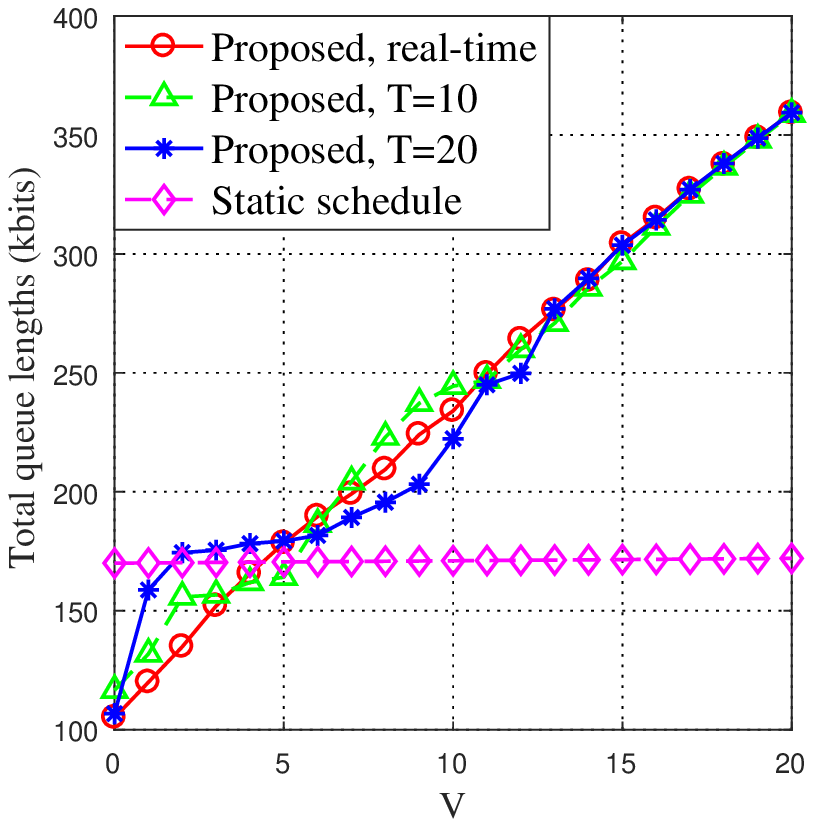}
	\end{minipage}
}
	\subfigure[Jain's fairness index vs. $V$.]{	
	\begin{minipage}[t]{0.31 \textwidth}
		\centering
		\includegraphics[width=2.5 in]{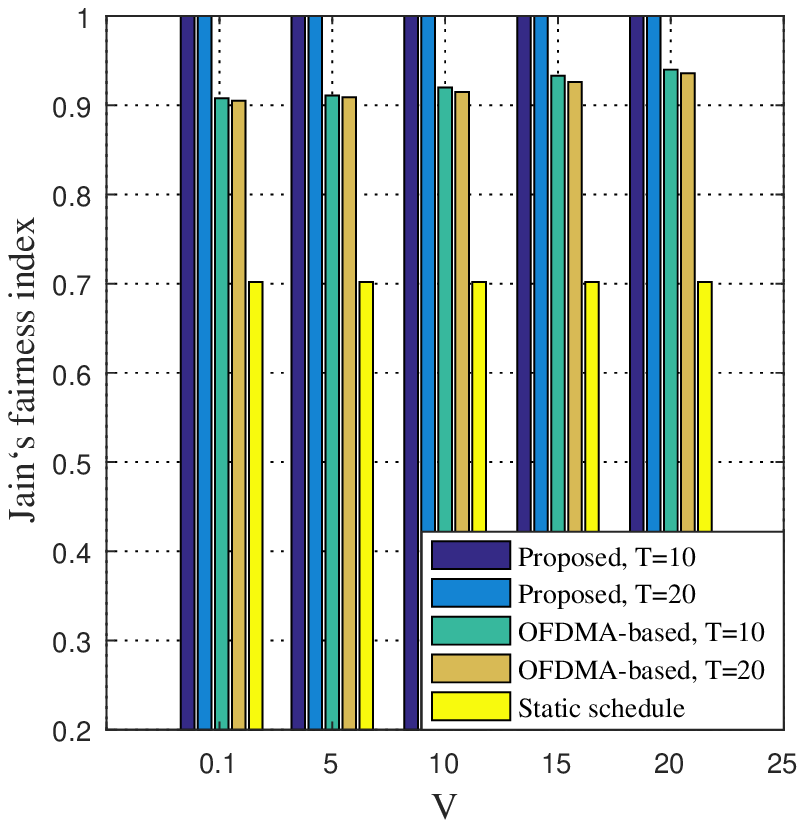}
	\end{minipage}
}
\caption{Impact of $V$ on average system throughput, total queue lengths and Jain's fairness index,  revealing the tradeoff between system utility and stability.}
\label{Impact_of_V_throughput_queue}
\end{figure*}

\begin{figure}[t] 
	\centering 
	\includegraphics[width=2.5in]{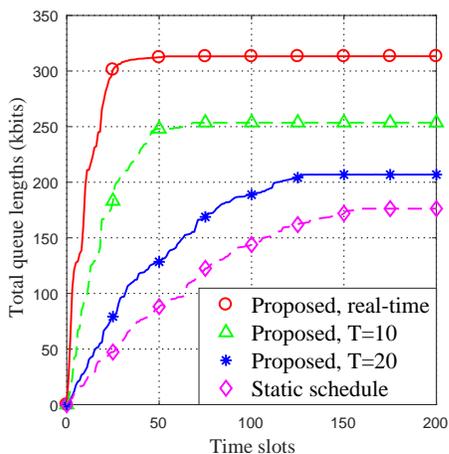} 
	\caption{The time variation of the total queue lengths. The proposed algorithm can accelerate the convergence of queues.} 
	\label{Total_Queue_Lengths_vs_time}
\end{figure}
\begin{figure}[t] 
	\centering 
	\includegraphics[width=2.5in]{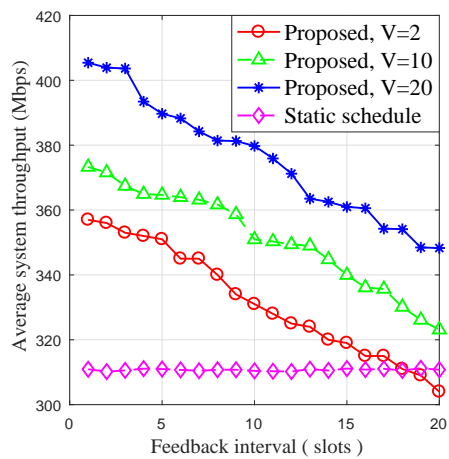} 
	\caption{Impact of feedback interval $T$ on average system throughput under different values of parameter $V$. 
	} 
	\label{system_utility_vs_T}
\end{figure}

\subsection{Simulation Settings}
	In this section, we evaluate the performance of the proposed online task offloading algorithm. We take the simulation parameters from state of the art \cite{Lyu2017Optimal,Wu2019Online,Zheng2021Achievable,Mao2017Stochastic,Cao2019Online,Nouri2020Dynamic,Qin2019Power}. Simulation parameter settings have been listed in Table \ref{simulation_para}, where $U_c[l,u]$ denotes a continuous uniform distribution within $[l,u]$, and $U_d[0,N]$ is a discrete integer uniform distribution from $0$ to $N$. For comparison, we simulate the following approaches.
\begin{enumerate}[1)]

\item \emph{Real-time Schedule:} The offloading strategies are operated instantaneously per time slot based on the complete device states $\boldsymbol{S}(t)$ as explained in Section \uppercase\expandafter{\romannumeral4}. The real-time scheduling mechanism provides an upper bound for the performance of our proposed algorithm. 

\item \emph{T-intervals Schedule:} The offloading strategies are conducted at each time slot according to the partial device states $\widehat{\boldsymbol{S}}(t)$, which is $T$ slots from the current time. 

\item \emph{SCA-based Schedule \cite{Ren2021An}:} The successive convex approximation (SCA) approach is utilized to allocate power by solving non-convex problem (29) iteratively. 
	
\item \emph{BnB-based Schedule \cite{Wang2019Multi}:} The branch and bound (BnB) method is adopted to obtain the optimal   offloading scheduling in the combinational optimization problem (23).

\item \emph{OFDMA-based Schedule \cite{Guo2014Throughput}:} The OFDMA-based approach is adopted to maximize throughput via subchannel and slot allocation, while considering long-term fairness. 

\item \emph{Static Schedule:} Offloading decisions and power allocation are made based on the static network environment, where the channel conditions and task arrivals are constants, and the  instantaneous system utility maximum is considered as the optimization objective.  
	
\end{enumerate}

\subsection{Result Analysis}
	Fig. \ref{system_utility_vs_arrival} compares the average system throughput of the four algorithms with the assumption of the overall device knowledge. The proposed approach performs the same with the BnB-based algorithm, and outperforms SCA-based algorithm and OFDMA-based algorithm in terms of the average system throughput. The key idea of SCA-based algorithm is to iteratively approximate the original non-convex problem (29) as multiple subproblems of locally optimal, which are a series of convex versions of the original problem. Therefore, it can obtain the suboptimal power allocation strategy. In the BnB-based algorithm, the set of candidate solutions of the offloading decisions is formed as a rooted tree, and each branch represents a subset of the candidate solutions. By successively checking branches of this tree, the BnB-based algorithm can find the globally optimal solutions to the discrete and combinational optimization problem (23). Nevertheless, BnB-based algorithm requires checking the upper and lower bounds for $2^M$ vertices by solving $2^{M+1}$ subproblems (31) at the worst case. Therefore, the BnB-based algorithm generally has higher time complexity than the proposed algorithm, especially when the number of devices is sufficiently large. The OFDMA-based algorithm suffers the throughput degradation when the number of devices is greater than $60$ in the considered network. This is mainly because that larger number of devices results in less subcarrier bandwidth allocated on each user in OFDMA, which is adverse to the system throughput.  

Fig. \ref{overhead_vs_device} plots the total signaling overhead of the proposed NOMA-based approach and the OFDMA-based approach, as the number of devices increases from $0$ to $200$. The total signaling overhead grows rapidly, exhibiting the power function relation between them, which is in line with the \emph{Little's Law} claimed in Section \uppercase\expandafter{\romannumeral3}. Every offloading and feedback from devices to the BS incurs signaling overhead in terms of time including waiting, transmission and computation delay. Therefore, slower transmission and computation rate leads to more tasks accumulated in the queues. From the curves, one can see that the total signaling overhead of the proposed algorithm is relatively small, and reduces the overhead to 50\% compared to the OFDMA-based counterparts, which corroborates that offloading via NOMA can dramatically reduce the latency. We can see in Figs. \ref{system_utility_vs_arrival} and \ref{overhead_vs_device} that the proposed algorithm outperforms the OFDMA-based schedule in large-scale networks, even under partial device knowledge. 

Fig. \ref{Impact_of_V_throughput_queue} (a) plots the curve of average system throughput as $V$ increases from $0.1$ to $20$. The average system throughput first increases  and stabilizes when $V \ge 10$, as revealed in \textbf{Theorem \ref{tradeoff}}. The average system throughput of the proposed $T=10/20$ offloading schedule is smaller than that of the real-time counterpart, and the difference between them diminishes with $V$ increasing, which corroborates the asymptotic optimality under partial knowledge as declared in \textbf{Theorem \ref{optimality_loss}}. Moreover, the proposed real-time and $T=10/20$ schedule outperform the static schedule in terms of the average system throughput, by up to 25\%, 21.8\% and  19.2\%, respectively, see $V=20$.

Fig. \ref{Impact_of_V_throughput_queue} (b) reveals that the total queue lengths increase linearly with $V$, especially when $V$ is greater than $5$. This trend shows that a larger $V$ helps increase the system throughput, but would also require longer system queue backlogs. The static schedule conducts offloading decisions and resource allocation for each device based on static system states. Therefore, the system throughput and queue lengths under static schedule are independent of $V$ and remain unchanged. 

Fig. \ref{Impact_of_V_throughput_queue} (c) shows the fairness against the control parameter $V$. Jain's index is utilized to quantify the fairness, and a larger Jain’s index corresponds to a fair resource allocation. Particularly, the fairest case appears when all the devices achieve the proportional fair throughput according to their weights, which makes the Jain’s index equal to one. It can be observed that the proposed algorithm maintains almost the same high fairness equal to one with different values of $V$, and is barely influenced by the feedback interval $T$. However, the fairness performance of the OFDMA-based algorithm is apparently affected by both of them, where a smaller $T$ and larger $V$ benefit to the fairness and vice versa. Associating Fig. \ref{Impact_of_V_throughput_queue} (a) to Fig. \ref{Impact_of_V_throughput_queue} (c), we can firmly conclude that there is an $\left[ \mathcal{O}\left(1/V \right) ,\mathcal{O} \left(V \right)  \right] $ tradeoff between system utility and queue backlogs, which verifies \textbf{Theorem \ref{tradeoff}}. 

Fig. \ref{Total_Queue_Lengths_vs_time} plots the time variation of the total queue lengths under different scheduling strategies. It can be observed that the total queue lengths  increase and then stabilize at different values. The stabilized queue lengths of the proposed $T=10/20$ schedule are shorter than that of the proposed real-time counterpart. The static schedule requires the shortest time for stabilization and implements the shortest queue length, but suffers from a 25\% discount on the system throughput compared to the real-time schedule as shown in Fig. \ref{Impact_of_V_throughput_queue} (a). Moreover, the real-time schedule is able to further speed up the stabilization and increase the backlog, since a larger system throughput is accompanied by longer queue lengths, which reflects the $[ \mathcal{O} (1/V) ,\mathcal{O} (V)] $ tradeoff. Compared to the real-time schedule, the proposed algorithm takes additional $150$ slots to stabilize, due to the larger interval for feedback. 

Fig. \ref{system_utility_vs_T} plots the average system throughput as the feedback interval $T$ increases from $1$ to $20$ slots. It is displayed intuitively that the average system throughput of the proposed algorithm  decreases linear with $T$, which is in line with the inequation (\ref{asymptotic_opt}). The device feedback takes up more time originally assigned for offloading, resulting in a reduction in the system throughput. Besides, the average system throughput also diminishes with the decrease of $V$, and a fairly small value of $V$ as well as an overlong feedback interval will cause a worse performance than the static schedule, which corresponds to $V \le 2$ and $T \ge 20$ under our simulation setup.

\section{Conclusion}

In this paper, we proposed an asymptotically optimal online offloading algorithm with complexity of $\mathcal{O} (M^{3/2})$ for long-term system utility maximum in NOMA-aided MEC networks while balancing system throughput and user fairness. By exploiting Lyapunov optimization, the offloading decisions and resource allocation were jointly considered at each time slot in complete and partial device knowledge situations. The proposed algorithm was proved to achieve asymptotic optimality,  presenting an $[ \mathcal{O}(1/V) ,\mathcal{O} (V)] $  tradeoff between system utility and stability. Simulations confirm our theoretical analysis, and reveal that our algorithm can significantly enhance the system utility to $85\%$ and $25\%$ compared with the OFDMA-based and static offloading approach, respectively. Besides, the signaling overhead is reduced by 50\% in comparison with the OFDMA-based algorithm. Feedback from partial devices can tackle the scalability problem in large-scale networks. Nevertheless, various device states, such as channel conditions and residual computation resource, have different impacts on the performance of NOMA. How to select devices for feedback to coordinate the scheduling leaves for future work.

\section*{Appendix A}
\section*{Proof of Lemma 1}

\label{proof_mean_rate_stable}
According to the dynamic of $Q_p(t) $ in (\ref{virtual_queue_of_power}), we naturally have the following inequality:
\begin{equation}
\label{dynamic_of_Q_p}
Q_p\left( t+1 \right)  \ge Q_p\left( t  \right) - p_{\text{ave}} + p_\text{tot}(t) .
\end{equation}

Summing (\ref{dynamic_of_Q_p}) in time $\left\{ {0,1, \cdots ,T} \right\}$ interval and taking expectations, we have
\begin{equation}
\label{sum_and_expectation}
\mathbb{E}\left[ {{Q_p}\left( T \right)} \right] \ge \sum\nolimits_{t = 0}^{T - 1} {\mathbb{E}\left[ {{p_\text{tot}}(t)} \right]}  - T{p_{\text{ave}}}.
\end{equation}

Dividing (\ref{sum_and_expectation}) by $T$ and taking $T \to \infty $, we have
\begin{equation}
\label{Dividing_T}
\mathop {\lim }\limits_{T \to \infty } \frac{1}{T}E\left[ {{Q_p}\left( T \right)} \right] \ge \mathop {\lim }\limits_{T \to \infty } \frac{1}{T}\sum\nolimits_{t = 0}^{T - 1} {E\left[ {{p_\text{tot}}(t)} \right]}  - {p_{\text{ave}}}.
\end{equation}

According to Jensen's inequality, we obtain $0 \le \left| {\mathbb{E}\left[ {{Q_p}(t)} \right]} \right| \le \mathbb{E}\left[ {\left| {{Q_p}(t)} \right|} \right]$. Based on Definition 2, if $Q_p\left(t \right) $ is mean rate stable, i.e., $\mathop {\lim }\nolimits_{T \to \infty } \frac{1}{T}\sum\nolimits_{t = 0}^{T - 1} {\mathbb{E}\left[ {{Q_p}(t)} \right]}  = 0$, the following equation holds:
\begin{align}
\label{left}
\mathop {\lim }\limits_{T \to \infty } \frac{1}{T} \mathbb{E}\left[ {{Q_p}\left( T \right)} \right] = 0.
\end{align}

Substitute (\ref{left}) into (\ref{Dividing_T}), and we have 
\begin{equation}
\mathop {\lim }\limits_{T \to \infty } \frac{1}{T}\sum\nolimits_{t = 0}^{T - 1} {\mathbb{E}\left[ {{p_\text{tot}}(t)} \right]}  \le {p_{\text{ave}}},
\end{equation}
which is consistent with (25f), and \emph{Lemma 1} is proved.

\section*{Appendix B}
\section*{Proof of Lemma 2}
\label{proof_dpp_up_bounded}
By squaring both sides of (\ref{localcomputing_queue}) and using the fact that 
$\left( \max \left(x,0\right)  \right)^2 \le x^2$, we have 
\begin{align}
& Q_{m,\text{loc}}^2(t+1) - Q_{m,\text{loc}}^2(t) \notag \\
&\le  A_m^2 (t) + \mu _m^2(t) - 2{Q_{m,\text{loc}}}(t)\left[ {{\mu _m}(t) -  {A_m}(t)} \right].
\end{align}

Adding up the above inequalities over $m = 1,2, \cdots, M$, we have
\begin{align}
& \sum\nolimits_{m = 1}^M {\left\{ {Q_{m,\text{loc}}^2(t+1) - Q_{m,\text{loc}}^2(t)} \right\}}  \notag \\
&\le M{\left[ {{A_{m,\max }}(t) + {\mu _{m,\max }}(t)} \right]^2} \notag \\
& - 2\sum\nolimits_{m = 1}^M {\left\{ {{Q_{m,\text{loc}}}(t)\left[ {{\mu _m}(t) - {A_m}(t)} \right]} \right\}}, \label{square_loc} 
\end{align} 
where ${{\mu _{m,\max }}(t)}$ is the maximum departure rate for local computing. Similarly, we obtain the following inequalities for $Q_{m,\text{off}}(t) $, $Q_{\text{BS}}(t)$ and $Q_{m,p}(t)$, respectively
\begin{align}
& \sum\nolimits_{m = 1}^M {\left\{ {Q_{m,\text{off}}^2(t+1) - Q_{m,\text{off}}^2(t)} \right\}} \notag \\
&  \le M{\left[ {{A_{m,\max }}(t) + {R_{m,\max }}(t)} \right]^2} \notag\\
&   - 2\sum\nolimits_{m = 1}^M {\left\{ {{Q_{m,\text{off}}}(t)\left[ {{R_{m  }}(t) - {A_{m  }}(t)} \right]} \right\}};   \label{square_off} 
\end{align}
\begin{align}
& Q_{\text{BS}}^2(t+1) - Q_{\text{BS}}^2(t) \notag \\
&\le  \{ {\sum\nolimits_{m = 1}^M {{R_{m,\max }}(t)}  + {\mu _{BS,\max  }}(t)} \} ^2  \notag \\
& - 2{Q_{\text{BS}}}(t) \{ {{\mu _\text{BS}}(t) - \sum\nolimits_{m = 1}^M {{R_{m  }}(t)} } \};
 \label{square_BS}  
\end{align}
\begin{align}
& \sum\nolimits_{m = 1}^M {\left\{ {Q_{m,p}^2(t+1) - Q_{m,p}^2(t)} \right\}}  \notag \\
&  \le M{\left[ {{p_{m,\text{tot},\max }}(t) + {p_{m,\text{ave}}}(t)} \right]^2} \notag\\
&  - 2\sum\nolimits_{m = 1}^M {\left\{ {{Q_{m,\text{off}}}(t)\left[ {{p_{m,\text{ave}}}(t) - {p_{m,\text{tot} }}(t)} \right]} \right\}}. \label{square_p} 
\end{align}

Summing over (\ref{square_loc})-(\ref{square_p}), dividing both sides by 2, rearranging terms, and we obtain the upper bound of the Lyapunov drift-plus-penalty function (\ref{up_bound}).
 
\section*{Appendix C}
\section*{Proof of Lemma 3}
\label{proof_convex}
According to (\ref{uplink_rate}), the uplink power allocation of device $m$ is
\begin{equation}
\label{p_m}
{p_m} ( t  ) =  ( {{2^{{R_m}(t)}} - 1}  ) ( {\sum\nolimits_{i = m + 1}^M {{p_i} ( t )}  + \frac{1}{{{g_{m,t}} }}}  ),\forall m \in \mathcal{M}.
\end{equation}

Specifically, 
\begin{equation}
\label{p_M}
{p_M}(t) = \frac{1}{{{g_{M,t}} }} ( {{2^{{R_M}(t)}} - 1} ).
\end{equation}

Substitute (\ref{p_M}) into (\ref{p_m}), we have 
\begin{equation}
\label{p_M_1}
{p_{M - 1}} ( t ) =  ( {{2^{{R_{M - 1}}(t)}} - 1} ) ( {{p_M}(t) + \frac{1}{{{g_{{M - 1},t}} }}} ).
\end{equation}

Adding (\ref{p_M}) and (\ref{p_M_1}) yields
\begin{align}
\label{delta_p_M}
& {p_M}(t) + {p_{M - 1}}(t) \notag \\
& = \frac{1}{g_{M,t}}  ( {2^{{R_M}(t) + {R_{M - 1}}(t)}} )+ ( {\frac{1}{{{g_{{M - 1},t}} }} - \frac{1}{{g_{M,t}} }}){2^{{R_{M - 1}}(t)}} \notag \\
&\quad - 1/ g_{{M - 1},t} .
\end{align}

Adopting the recursive method, the total uplink power of all devices can be derived as 
\begin{align}
\label{total_uplink_power}
\sum\limits_{m = 1}^M {{p_m}(t)}  = 
\sum\limits_{m = 1}^M {( \frac{1}{g_{m,t}} - \frac{1}{g_{m + 1,t}} )} {2^{\sum\nolimits_{i = 1}^m {{R_i}(t)} }} - \frac{1}{{g_{1,t}}}.
\end{align}

Note that $1/{g_{M + 1,t}} \buildrel \Delta \over = 0$. With ${\nu _{m,t}} = \sum\nolimits_{i = 1}^m {{R_i}(t)} ,\forall m \in M $, we have
\begin{equation}
\label{R_m}
R_m(t) = {\nu _{m,t}} - {\nu _{m - 1,t}} , \ \forall m \in M.
\end{equation}

Substituting (\ref{total_uplink_power}) and (\ref{R_m}) into problem (\ref{power_allocation}), we obtain the equivalent formulation as represented in problem (\ref{reformulated}).

\section*{Appendix D}
\section*{Proof of Theorem 2}
\label{proof_tradeoff}
To prove the upper bounds of average system utility and queue length, Lemma \ref{lemma_in_appendix} is introduced first:

\begin{lemma}
\label{lemma_in_appendix}
For arbitrary arrival rates and random channel conditions, there exists a stationary and randomized policy, which determines a feasible operation, $\boldsymbol{x_t}$, independent of current actual queues and virtual queues among different time slots and satisfies all the instantaneous constraints. Then for any $\varepsilon $, $0 \le \varepsilon  \le {\varepsilon _{\max }}\left( {\rm{x}} \right)$, and $\delta > 0 $, the following steady state is satisfied:
\begin{align}
& \mathbb{E} \left[ {R_m^t\left( {{{\rm{x}}_{{\rm{t}}}}} \right)} \right] \ge {\rho _{m,t}}{A_m}(t) + \varepsilon ; \label{E1}    \\
& \mathbb{E} \left[ {\mu _m^t\left( {{{\rm{x}}_{{\rm{t}}}}} \right)} \right] \ge \left( {1 - {\rho _{m,t}}} \right){A_m}(t) + \varepsilon; \label{E2}  \\
& \mathbb{E} \left[ {\mu _{\text{BS}}^t\left( {{{\rm{x}}_{{\rm{t}}}}} \right)} \right] \ge \sum\nolimits_{m = 1}^M {{R_m}(t)}  + \varepsilon; \label{E3} \\
& \mathbb{E} \left[ {p_{m,\text{tot}}^t\left( {{{\rm{x}}_{{\rm{t}}}}} \right)} \right] \le {p_{\text{ave}}} - \delta. \label{E4} 
\end{align}
\end{lemma}

The proof of Lemma \ref{lemma_in_appendix} can refer to Theorem 4.5 in \cite{Neely2010Stochastic}, which is omitted for brevity. Substituting (\ref{E1})-(\ref{E4}) into (\ref{up_bound}) and taking the limit $\delta  \to 0$, we have 
\begin{align}
& {\Delta _V}\left( {{ \boldsymbol{\Theta} (t) }} \right) \le B - V\mathbb{E}\left[ {\left. {{{  U}^*}(t)} \right|{ \boldsymbol{\Theta} (t) }} \right]\notag \\ 
& - \varepsilon \sum\nolimits_{m = 1}^M {  \left[  {Q_{m,\text{off}}}(t) + {Q_{m,\text{loc}}}(t) \right]   }  - \varepsilon {Q_{\text{BS}}}(t). \label{E5} 
\end{align}

By using telescoping sums over $t \in \left\lbrace 0,1,\cdots,T-1 \right\rbrace $ and taking the iterated expectation in (\ref{E5}), we get 
\begin{align}
& \mathbb{E}\left\{ {L\left( {{ \boldsymbol{\Theta} \left(T\right) }} \right)} \right\} - \mathbb{E}\left\{ {L\left( {{ \boldsymbol{\Theta} \left(0\right) }} \right)} \right\} - V\mathbb{E}\left[ {U(t)} \right] \notag \\
& \le BT - VT{{  U}^*}(t) \notag\\
&  - \varepsilon \sum\limits_{t = 0}^{T - 1} {\left\{ {\sum\limits_{m = 1}^M {\left[ {{Q_{m,\text{off}}}(t) + {Q_{m,\text{loc}}}(t)} \right]}  + {Q_{\text{BS}}}(t)} \right\}}. \label{E6} 
\end{align}

Dividing (\ref{E6}) by $VT$, negelecting the  nonnegative terms $Q_{m,\text{loc}}(t)$,  $Q_{m,\text{off}}(t)$, $Q_{\text{BS}}(t)$ and $\mathbb{E}\left\{ {L\left( {{ \boldsymbol{\Theta} \left(T\right) }} \right)} \right\}$, and rearranging terms, we obtain  
\begin{align}
& \frac{1}{T}\mathbb{E}\left[ {U(t)} \right] \ge {{  U}^*}(t) - \frac{B}{V} - \frac{{\mathbb{E}\left\{ {L\left( {{\boldsymbol{\Theta} \left( 0\right) }} \right)} \right\}}}{{VT}}.  \label{E7} 
\end{align}

Dividing (\ref{E6}) by $V\varepsilon $, we have
\begin{align}
& \frac{1}{T}\sum\limits_{t = 0}^{T - 1} {\left\{ {\sum\limits_{m = 1}^M {\left[ {{Q_{m,\text{off}}}(t) + {Q_{m,\text{loc}}}(t)} \right]}  + {Q_{\text{BS}}}(t)} \right\}}  \notag \\
& \le \frac{B}{\varepsilon } - \frac{{V\left\{ {{{  U}^*}(t) - \frac{1}{T}\sum\limits_{t = 0}^{T - 1} {\mathbb{E}\left[ {U(t)} \right]} } \right\}}}{\varepsilon } + \frac{{\mathbb{E}\left\{ {L\left( {{\boldsymbol{\Theta} \left( 0\right)}} \right)} \right\}}}{{\varepsilon T}} .\label{E8}  
\end{align}

Taking the limit as $T \to \infty $ in (\ref{E7}) and (\ref{E8}), yields
\begin{align}
{U ^*\left(t \right) }  - {{\overline U }^*\left( t\right) } \le B/V ;
\end{align}
\begin{align}
& \mathop {\lim }\limits_{T \to \infty } \sup \sum\nolimits_{m = 1}^M {\mathbb{E}\left\{ {{Q_{m,\text{loc}}}(t) + {Q_{m,\text{off}}}(t)} \right\}}  + {Q_{\text{BS}}}(t) \notag\\
& \qquad  \qquad \qquad  \qquad \le \frac{B}{\varepsilon } - \frac{{V\left[ {{U^*}(t) - {{\bar U}^*}(t)} \right]}}{\varepsilon } .
\end{align}

\section*{Appendix E}
\section*{Proof of Lemma 4}
\label{proof_difference_of_queue}
	For two adjacent slot $t$ and $t+1$, the difference of the queue lengths is bounded by the maximum difference between the task arrival and the data departure, as captured in the first and second terms of $\delta _{\{i\}}$ respectively. Considering the worst case, where the feedback interval is $T$, we have $ |{{\widehat Q}_{\{i\}}(t) - Q_{\{i\}}(t)} | = | Q_{\{i\}}(t- T\Delta t)-Q_{\{i\}}(t) |\le  T {\delta _{\{i\}}}$, where the first inequality is derived from reusing absolute value inequality $|x+y| \le |x|+|y|$. 

\section*{Appendix F}
\section*{Proof of Theorem 3}
\label{proof_difference_of_approximate}
	Regarding ${H_1}( \boldsymbol{{\widehat \rho }}_t)$, we can rewrite it as 
	\begin{equation}
	\label{app_H1}
	\begin{array}{l}
	{H_1}( \boldsymbol{{\widehat \rho }}_t )  
	= \sum\limits_{m = 1}^M \{ \widehat \alpha_m(t){{\widehat \rho }_{m,t}} 
	+  {Q}_{m,loc}(t) A_m(t) \\
	\qquad \qquad \qquad + \left( {{\alpha _m}(t) - {{\widehat \alpha }_m}(t)} \right)\widehat \rho_{m,t}.
	\end{array}
	\end{equation}
	
	By defining ${\alpha _m}(t) \buildrel \Delta \over = [ {Q_{m,\text{off}}(t) - Q_{m,\text{loc}}(t)}]{A_m}(t)$, we have 
	\begin{align*}
	{\widehat \alpha _m}(t) - {\alpha _m}(t)  
	& =  \{ [ {\widehat Q}_{m,\text{off}}(t) - Q_{m,\text{off}}(t)] \\
	& - [ \widehat Q_{m,\text{loc}}(t) - Q_{m,\text{loc}}(t) ] \} {A_m}(t).
	\end{align*}
	
	Exploiting (\ref{delta_Q}) in \textbf{Lemma \ref{approximate_queue_bound}}, we have 
	\begin{equation}
	\label{alpha}
	| {{{\widehat \alpha }_m}(t) - {\alpha _m}(t)} | \le \max \left\{ {{\delta _\text{off}},{\delta _\text{loc}}} \right\} T {A_m}(t).
	\end{equation}
	
	Recall that $\boldsymbol{{\widehat \rho }}_t$ is the optimal offloading decision under partial knowledge, and we have
	\begin{equation}
	\begin{array}{l}
	\sum\limits_{m = 1}^M {{{\widehat \alpha }_m}(t){{\widehat \rho }_{m,t}}} 
	\le {H_1}\left( \boldsymbol{\rho}_t  \right) + \sum\limits_{m = 1}^M {\left( {{{\widehat \alpha }_m}(t) - {\alpha _m}(t)} \right){\rho _{m,t}}}. \notag
	\end{array}
	\end{equation}

	Substituting (\ref{alpha}) into the subtraction of (39) and (40), we obtain
	\begin{equation}
	\label{diff_H1}
	\begin{array}{l}
	{H_1}( \widehat {\boldsymbol{ \rho }}_t ){\rm{ - }}{H_1}( \boldsymbol{\rho}_t ) \\
	\le \sum\limits_{m = 1}^M  \{ [ \widehat \alpha _m(t)\widehat \rho_{m,t} + ( {{\alpha _m}(t) - {{\widehat \alpha }_m}(t)} ){{\widehat \rho }_{m,t}}] \\
	\qquad \quad + \left( {{{\widehat \alpha }_m}(t) - {\alpha _m}(t)} \right){\rho _{m,t}}  - \widehat \alpha _m(t){\widehat \rho }_{m,t} \} \\
	\le \sum\limits_{m = 1}^M {| {{{\widehat \alpha }_m}(t) - {\alpha _m}(t)} |} \left( {{{\widehat \rho }_{m,t}} + {\rho _{m,t}}} \right)\\
	\le 2T {A_m}(t)\sum\limits_{m = 1}^M {\max \{ {{\delta _\text{off}},{\delta _\text{loc}}} \}} .
	\end{array}
	\end{equation}
	
	Likewise, we can evaluate the deficiency of $H_2(\boldsymbol{f}(t) ) $ and $ {H_3} ( \boldsymbol{\nu}(t))$ with, respectively 
	\begin{equation}
	\label{diff_H2}
	\begin{array}{l}
	{H_2}(\widehat{\boldsymbol{f}}(t) )  - {H_2}(\boldsymbol{f}(t) )  \\
	\le \sum\limits_{m = 1}^M {2{f_{\max }} [ {{\varepsilon T}\delta _\text{loc}/C_m(t) + f_{\max }^2{\kappa _m}T{\delta _p}} ] }, 
	\end{array}
	\end{equation}
	\begin{equation}
	\label{diff_H3}
	\begin{array}{l}
	{H_3} (\widehat{ \boldsymbol{\nu} }(t) )- {H_3}\left( \boldsymbol{\nu} \left( t\right)  \right)  \\ 
	\le \sum\limits_{m = 1}^M {\left\{ {2T{\delta _{m,\text{off}}}\ln \left( {M{g_{1,t }}{p_\text{max}} + 1} \right) + 2T{p_\text{max}}{\delta _p}} \right\}}  \\
	+ 2T{\delta _\text{BS}}\ln  ( p_{M,t } +  {1}/Mg_{1,t } )  (1/g_{M,t} - 1/g_{M - 1,t} )^{ - 1}  \\
	+ V\sum\limits_{m = 1}^M \log \{ 1+ \ln [ (M{g_{1,t}}{p_\text{max}})^2 - e (M{g_{1,t}}{p_\text{max}})^{-1} ] \}. 
	\end{array}
	\end{equation}
	
	Constant $J$ is obtained by adding (\ref{diff_H1}), (\ref{diff_H2})  and (\ref{diff_H3}), and we prove the theorem.

\bibliographystyle{IEEEtran} 
\bibliography{IEEEabrv,ref} 

\begin{IEEEbiography}[{\includegraphics[width=1 in,height =1.3in,clip,keepaspectratio]{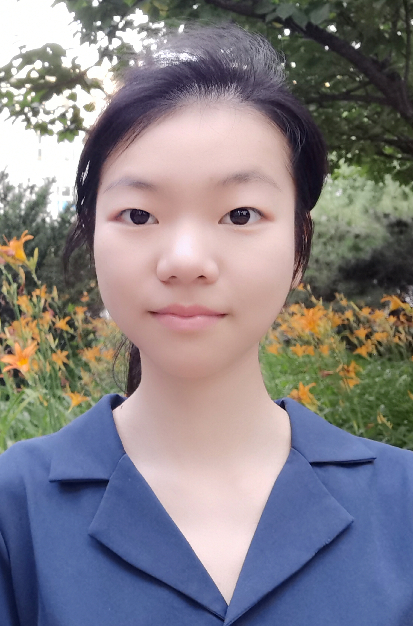}}]{Meihui Hua}
	received her B.E. degree in communication engineering from Beijing University of Posts and Telecommunications (BUPT), China, in 2019. She is currently working toward the Master degree in the State Key Laboratory of Networking and Switching Technology at BUPT. Her research interests include non-orthogonal multiple access (NOMA), mobile edge computing (MEC), cell-free massive MIMO, and resource allocation in wireless networks.
\end{IEEEbiography}

\begin{IEEEbiography}[{\includegraphics[width=1in,height=1.3in,clip,keepaspectratio]{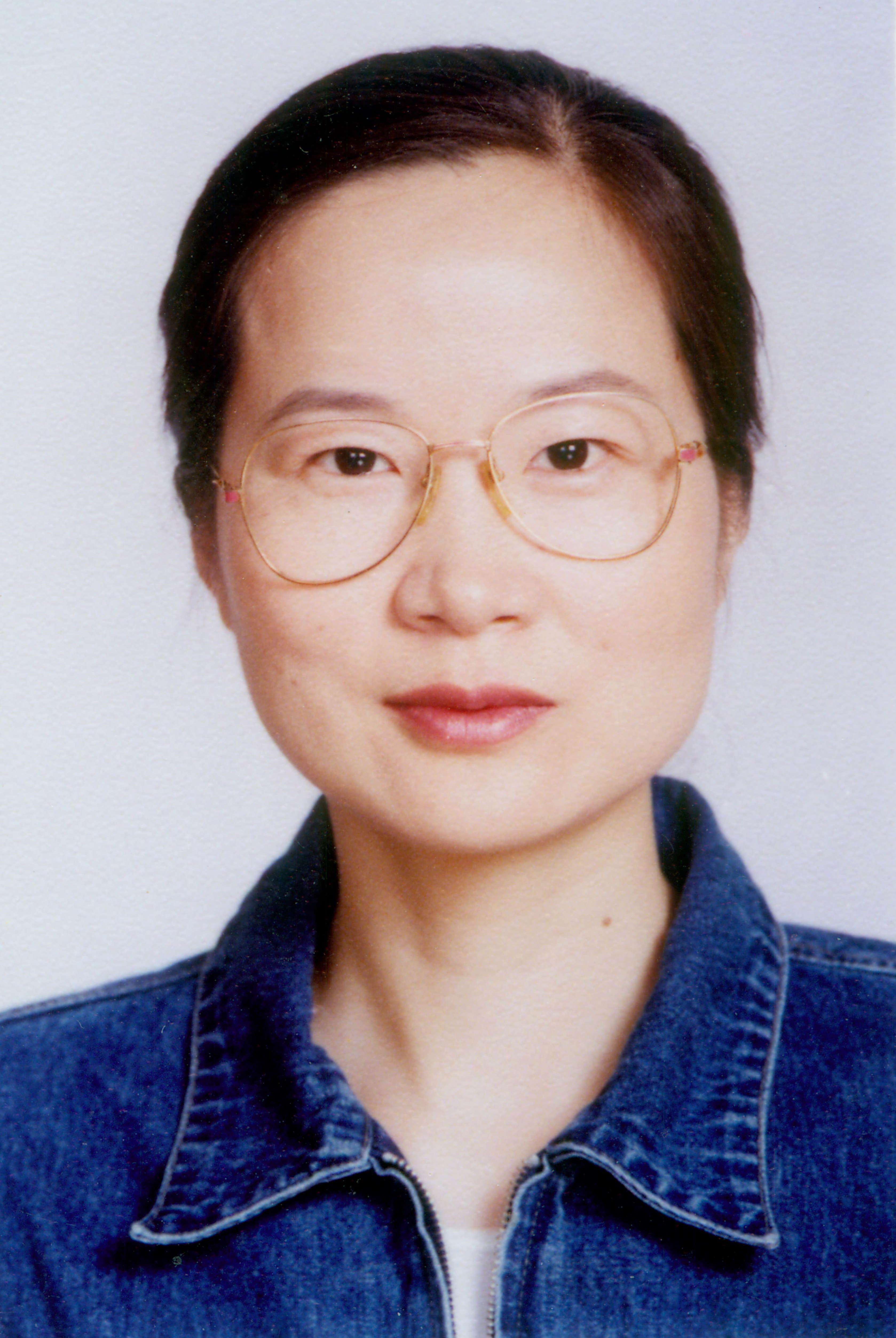}}]{Hui Tian}
	(Senior Member, IEEE) received the M.S. degree in Micro-electronics and the Ph.D. degree in circuits and systems both from Beijing University of Posts and Telecommunications (BUPT), China, in 1992 and 2003, respectively.
	Currently, she is a professor at BUPT, the Network Information Processing Research Center director of the State Key Laboratory of Networking and Switching Technology.
	Her current research interests mainly include radio resource management, cross layer optimization, M2M, cooperative communication, mobile social network, mobile edge computing, and edge intelligence.
\end{IEEEbiography}

\begin{IEEEbiography}[{\includegraphics[width=1in,height =1.3in,clip,keepaspectratio]{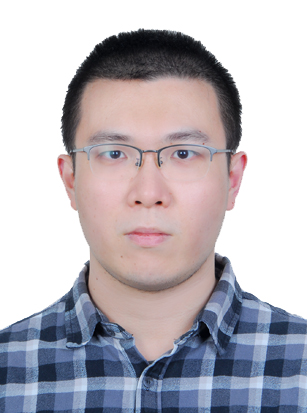}}]{Xinchen Lyu}
	(Member, IEEE) received the B.E. degree from the Beijing University of Posts and Telecommunications (BUPT) in 2014, and the dual Ph.D. degree from BUPT and the University of Technology Sydney in 2019. He is currently an Associate Researcher with BUPT. His research interests include stochastic optimization and machine learning with applications to fog computing, edge caching, software-defined networking, and resource management.
\end{IEEEbiography}

\begin{IEEEbiography}[{\includegraphics[width=1 in,height =1.3in,clip,keepaspectratio]{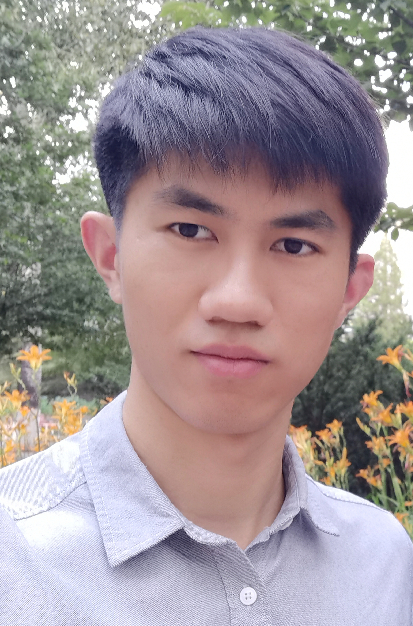}}]{Wanli Ni}
	(Student Member, IEEE) received his B.E. degree in communication engineering from Beijing University of Posts and Telecommunications (BUPT), China, in 2018. He is currently working toward the Ph.D. degree in the State Key Laboratory of Networking and Switching Technology at BUPT.
	He serves as a reviewer for the IEEE TRANSACTIONS ON WIRELESS COMMUNICATIONS, the IEEE TRANSACTIONS ON VEHICULAR TECHNOLOGY, and IEEE COMMUNICATIONS LETTERS.
	His research interests include reconfigurable intelligent surface (RIS), non-orthogonal multiple access (NOMA), mobile edge computing (MEC), over-the-air federated learning, resource allocation and machine learning in wireless networks. He was a recipient of the IEEE SAGC Best Paper Award in 2020.
\end{IEEEbiography}

\begin{IEEEbiography}[{\includegraphics[width=1 in,height =1.3in,clip,keepaspectratio]{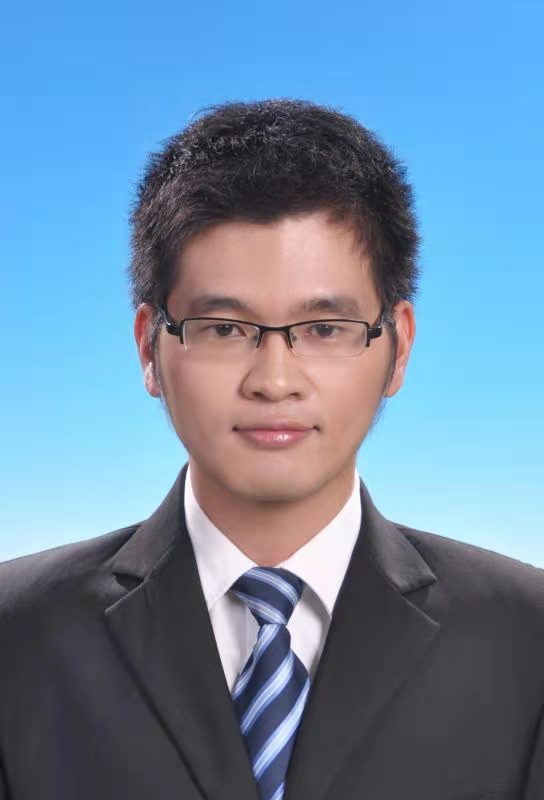}}]{Gaofeng Nie}
	(Member, IEEE) received the B.S. degree in communications engineering and the Ph.D. degree in telecommunications and information system from the Beijing University of Posts and Telecommunications (BUPT), in 2010 and 2016, respectively. He is currently a Lecturer with BUPT. His research interests include SDN over wireless networks and key technologies in 5G/B5G wireless networks.
\end{IEEEbiography}

\end{document}